\documentclass{PoS}

\usepackage{url}
\usepackage{amsmath}
\usepackage{amssymb}
\usepackage{graphicx}
\usepackage{bm}
\usepackage{epsfig}
\usepackage{amsfonts}
\usepackage{color}
\usepackage{mathtools}
\usepackage[section]{placeins}
\usepackage{wrapfig}
 \usepackage{notoccite}
 \usepackage[numbers]{natbib}

\title{In-medium heavy quarkonium from lattice NRQCD}

\ShortTitle{In-medium heavy quarkonium from lattice NRQCD}

\author{Seyong Kim\\
        Department of Physics, Sejong University, Seoul 143-747, Korea\\
        E-mail: \email{skim@sejong.ac.kr}}

\author{Peter Petreczky\\
        Physics Department, Brookhaven National Laboratory, Upton, NY 11973, USA\\
        E-mail: \email{petreczk@quark.phy.bnl.gov}}

\author{\speaker{Alexander Rothkopf}\\
        Faculty of Science and Technology, University of Stavanger, NO-4036 Stavanger, Norway\\
        E-mail: \email{alexander.rothkopf@uis.no}}

\abstract{We present the final results from a multi-year study of the in-medium spectral properties of heavy quarkonium bound states on the lattice. In this work we combine high statistics $N_f=2+1$ ensembles from the HotQCD collaboration with the effective theory NRQCD and improved Bayesian spectral reconstruction methods. We corroborate with high precision the hierarchical in-medium modification of quarkonium states with respect to their vacuum binding energy and provide updated values on melting temperatures. In particular we are able to understand previous disagreements between different Bayesian methods as resulting from underestimated systematic uncertainties. 
The main quantitative result is a robust determination of the in-medium mass shifts of quarkonium ground states, which we find are negative, consistent with the behavior observed in strongly coupled pNRQCD potential based computations.}

\FullConference{XIII Quark Confinement and the Hadron Spectrum - Confinement2018\\
		31 July - 6 August 2018\\
		Maynooth University, Ireland}

\begin{document}

\section{Motivation and methods overview}

Heavy quarkonium, the bound states of a heavy quark and its antiquark, have matured into a versatile precision probe in the experimental study of relativistic heavy-ion collisions (HIC) at RHIC and LHC. The presence of two different flavors allow us to probe different regimes of the time evolution of the collisions. Bottomonium shows excited state suppression consistent with a non-equilibrium probe sampling the full dynamical evolution of the quark-gluon-plasma (QGP) \cite{Krouppa:2017jlg}. On the other hand the ALICE collaboration observed that the $J/\Psi$ particle at the LHC shows a finite elliptic flow \cite{Acharya:2017tgv}, which indicates a partial kinetic equilibration with the bulk. This in turn implies a loss of memory of the initial conditions, positioning it as probe of the late stages. What we need to keep in mind however is that experiments measure the decay of vacuum $Q\bar Q$ states long after the QGP has ceased to exist. I.e. all of their medium interaction needs to be translated into vacuum states at hadronization, a process, which is far from understood from 1st principles. 

In this study \cite{Kim:2018yhk} we set out to explore one facet of this challenging physics puzzle, fully thermalized heavy quarkonium in a static medium. In order to capture the non-perturbative physics of the quarkonium in a QGP close the the phenomenologically relevant crossover transition, we turn to lattice QCD. In standard lattice formulations the light medium d.o.f's share the same spacetime grid as the heavy flavors (c,b) and one needs to adopt very fine lattice spacings, making simulations too costly. Instead this separation of scales $T/m_Q \ll 1$ and $\Lambda_{\rm QCD}/T\ll 1$ presents an advantage to deploy an effective non-relativistic field theory (NRQCD) for the heavy quarks. Lattice NRQCD \cite{Lepage:1992tx} is a well established tool, based on a systematic expansion of the QCD Lagrangian in powers of $1/m_Qa$. It is directly applicable at finite temperature and our study utilizes a Lagrangian up to order ${\cal O}(v^4)$, i.e. ${\cal O}(1/(m_Qa)^3)$ and leading order Wilson coefficients (including Tadpole improvement).

The medium d.o.f. are captured by realistic and high statistics lattice simulations by the HotQCD collaboration \cite{Bazavov:2014pvz,Bazavov:2011nk}. The pion mass is $m_\pi=161$MeV and the accessible temperature range $T\in[140-407]$MeV is explored by a change of the lattice spacing ($N_\tau=12$). For calibration purposes $T=0$ lattices are available ($N_\tau=32-64$). Compared to our previous study \cite{Kim:2014iga} we here consider not only bottom but also charm quarks and have collected much larger statistics. The parameter relevant for NRQCD is $m_ba\in[2.759-1.559]$, which is acceptable over the full $T$ range. For charm we restrict ourselves to $T\in[140-251]$MeV, where $m_c a\in[0.757 – 0.427]$. For a stable time evolution we choose the Lepage discretization parameter $n_{\rm b}=4$ and $n_{\rm c}=8$.

\begin{wrapfigure}{l}{0.31\textwidth}\vspace{-0.5cm}
\centering
 \includegraphics[scale=0.32]{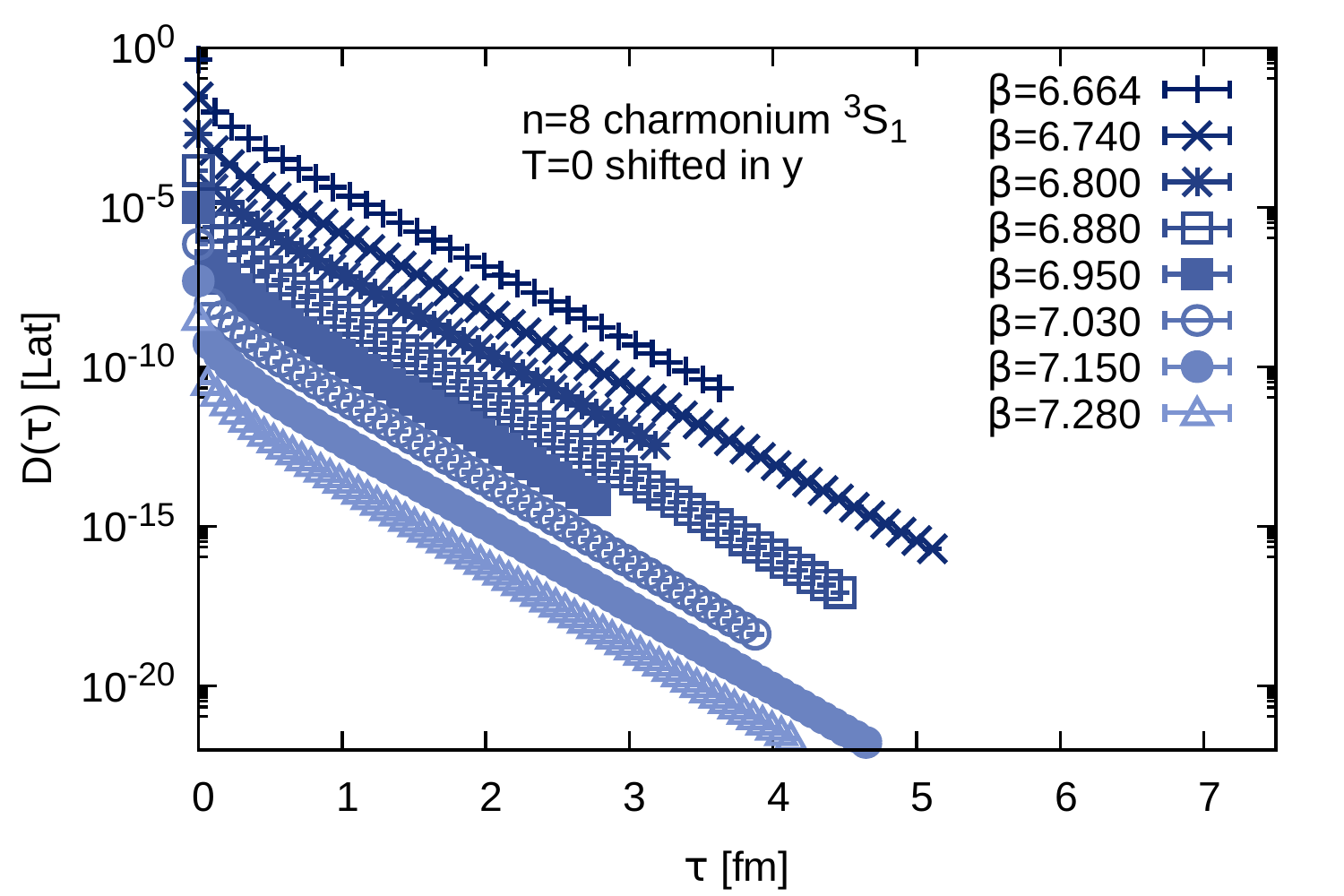}
 \includegraphics[scale=0.32]{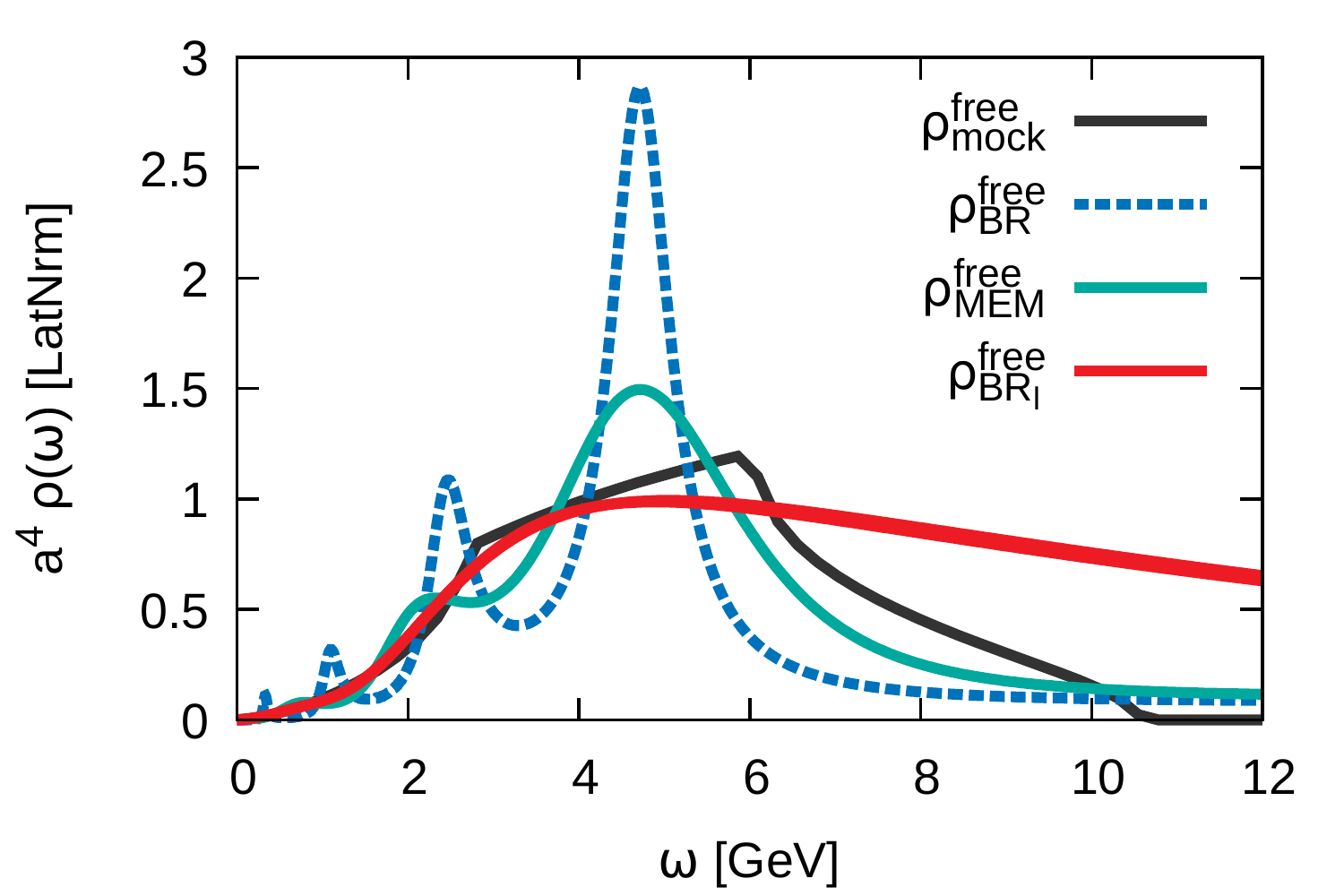} \vspace{-0.8cm}
 \caption{ (top) $^3S_1$ $c\bar c$ $T=0$ NRQCD correlator (bottom) mock reconstructing the non-interacting P-wave spectra} \label{Fig:T0Prep} \vspace{2.8cm}
\end{wrapfigure}

In NRQCD one computes the propagator of a single heavy quark in the background of the medium fields and subsequently combines two of these into a heavy quarkonium correlation function. Inserting appropriate vertex operators, the correlator may be projected into a channel with desired quantum numbers. At the end we arrive at Euclidean correlation functions, a $T=0$ example is plotted in Fig.\ref{Fig:T0Prep} and shows a clear exponential falloff related to a well defined ground state. For $b\bar b$ we compute $N^{T=0}_{\rm meas}=400$ and $N^{T>0}_{\rm meas}=1-4\times10^3$ correlators, for $c\bar c$ $N^{T=0}_{\rm meas}=N^{T>0}_{\rm meas}=400$.

While it is possible to extract vital insight on the overall in-medium modification of quarkonium from correlators alone, the in-medium properties of individual states may only be learned when we have access to spectral functions. To this end in NRQCD we need to invert a Laplace transform, a classic ill-posed problem, which we approach using Bayesian inference. Two improvements are part of this study: One the one hand we incorporate both the Euclidean correlator, as well as its Fourier transform in Matsubara frequencies in the reconstruction. As the inversion process is non-linear this leads to a more stable reconstruction at large frequencies. On the other hand we deploy besides two standard methods, the Maximum Entropy Method (MEM) \cite{Asakawa:2000tr} and the BR method \cite{Burnier:2013nla} a novel smooth variant of the BR \cite{Fischer:2017kbq}  
\begin{align}
 \nonumber &S_{\rm BR}=\alpha\int d\omega\Big(1-\frac{\rho}{m}+{\rm log}\big[ \frac{\rho}{m}\big]\Big),\\
 &S_{\rm BR_\ell}=\alpha\int d\omega\Big(\big(\frac{\partial\rho}{\partial \omega}\big)^2+1-\frac{\rho}{m}+{\rm log}\big[ \frac{\rho}{m}\big]\Big).
\end{align}
The standard BR method with regulator $S_{\rm BR}$ has been shown to resolve narrow peaks with high accuracy but may suffer from ringing if the number of input data is small. In order to combat ringing we introduce an additional derivative term in the regulator $S_{\rm BR_\ell}$, which penalizes arc length. \begin{wrapfigure}{r}{0.30\textwidth}\vspace{-1.0cm}
  \begin{center}
   \includegraphics[scale=0.30,trim=0cm 0.99cm 0cm 0cm, clip=true]{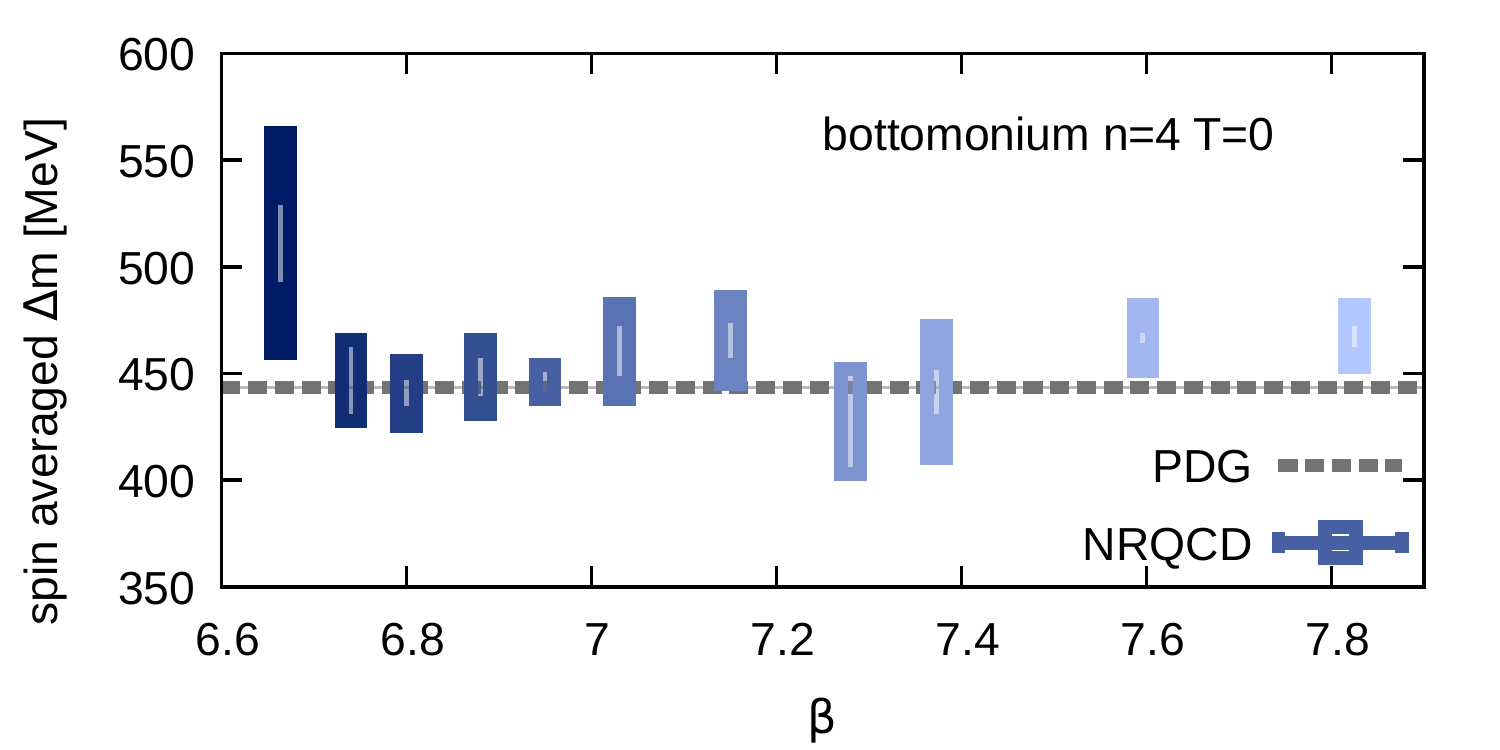}
   \includegraphics[scale=0.30,trim=0cm 0.99cm 0cm 0cm, clip=true]{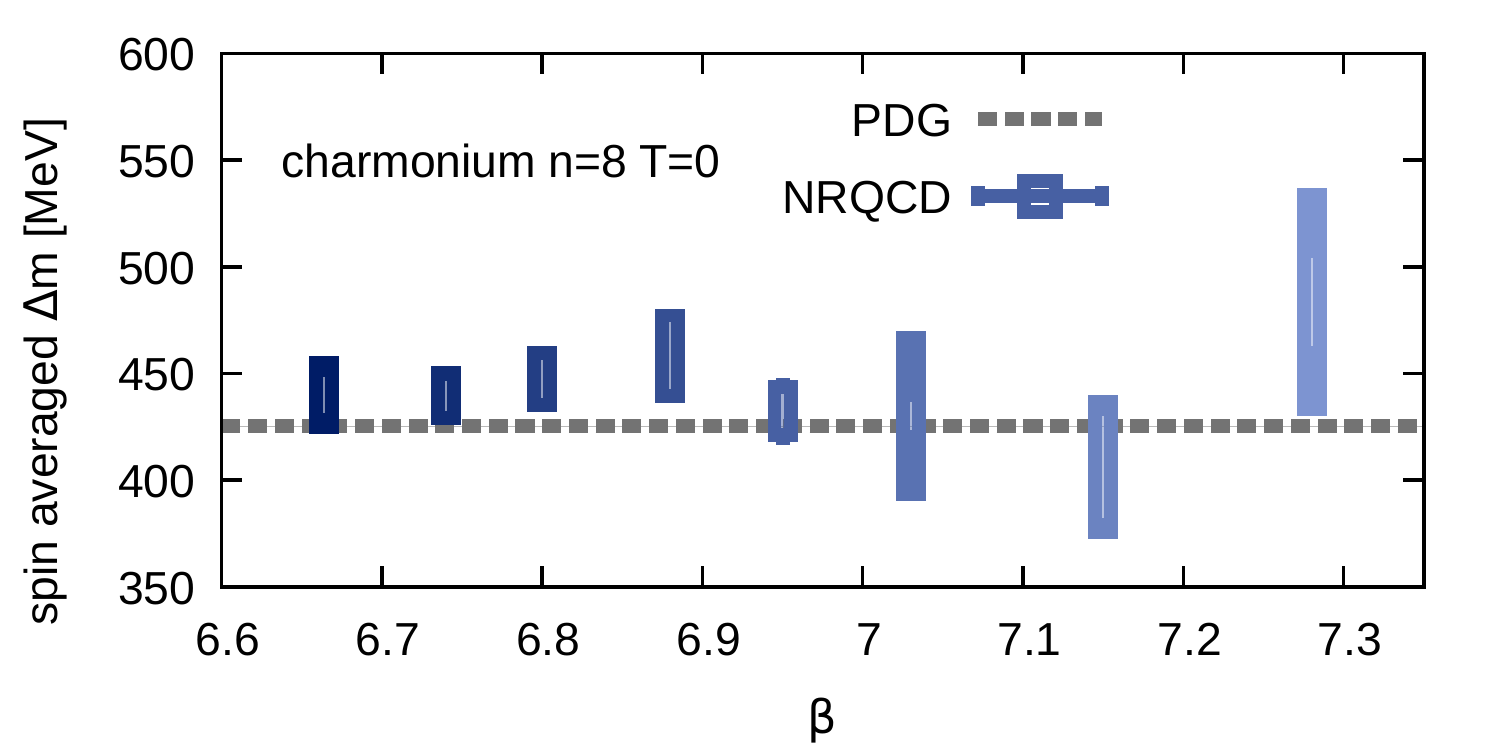}
   \includegraphics[scale=0.30,trim=0cm 0.99cm 0cm 0cm, clip=true]{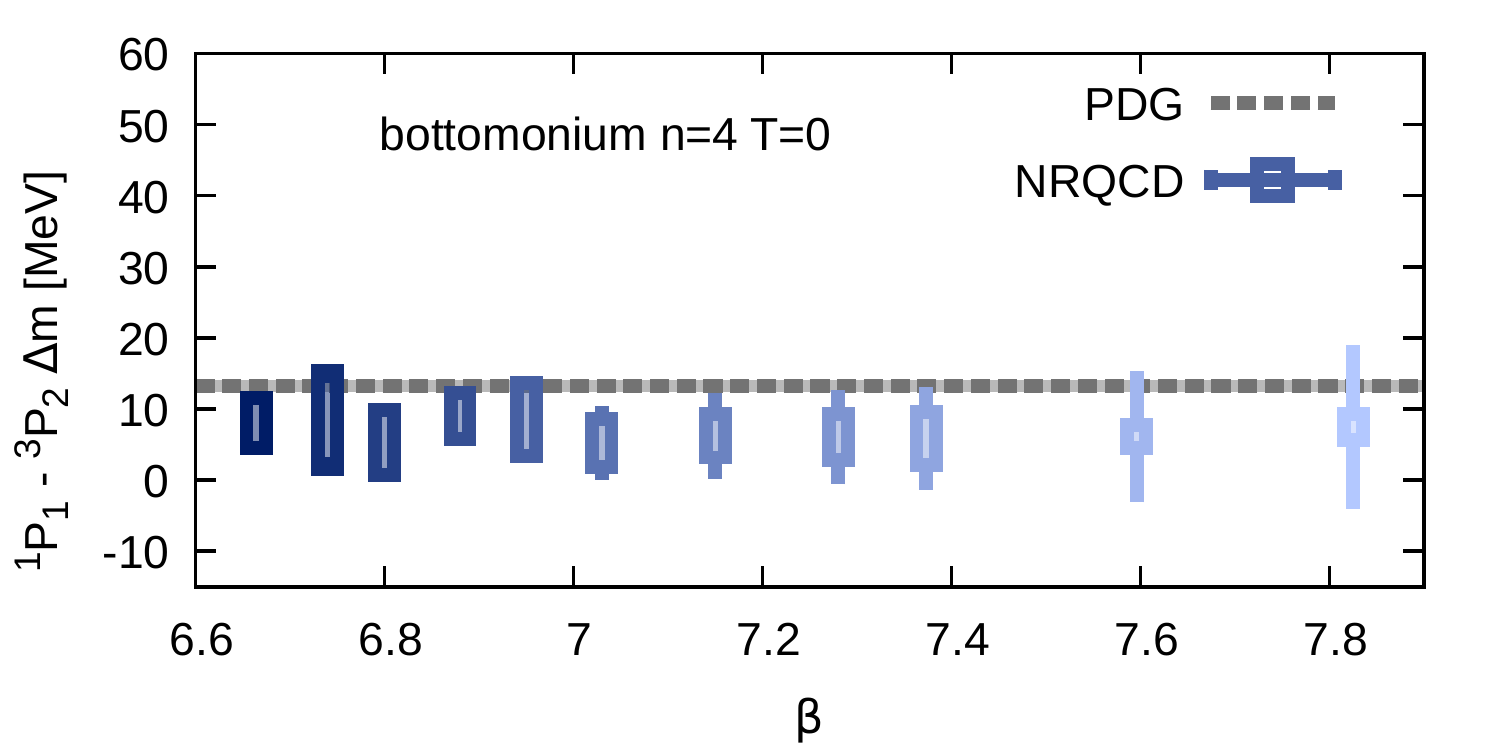}
   \includegraphics[scale=0.30,trim=0cm 0.99cm 0cm 0cm, clip=true]{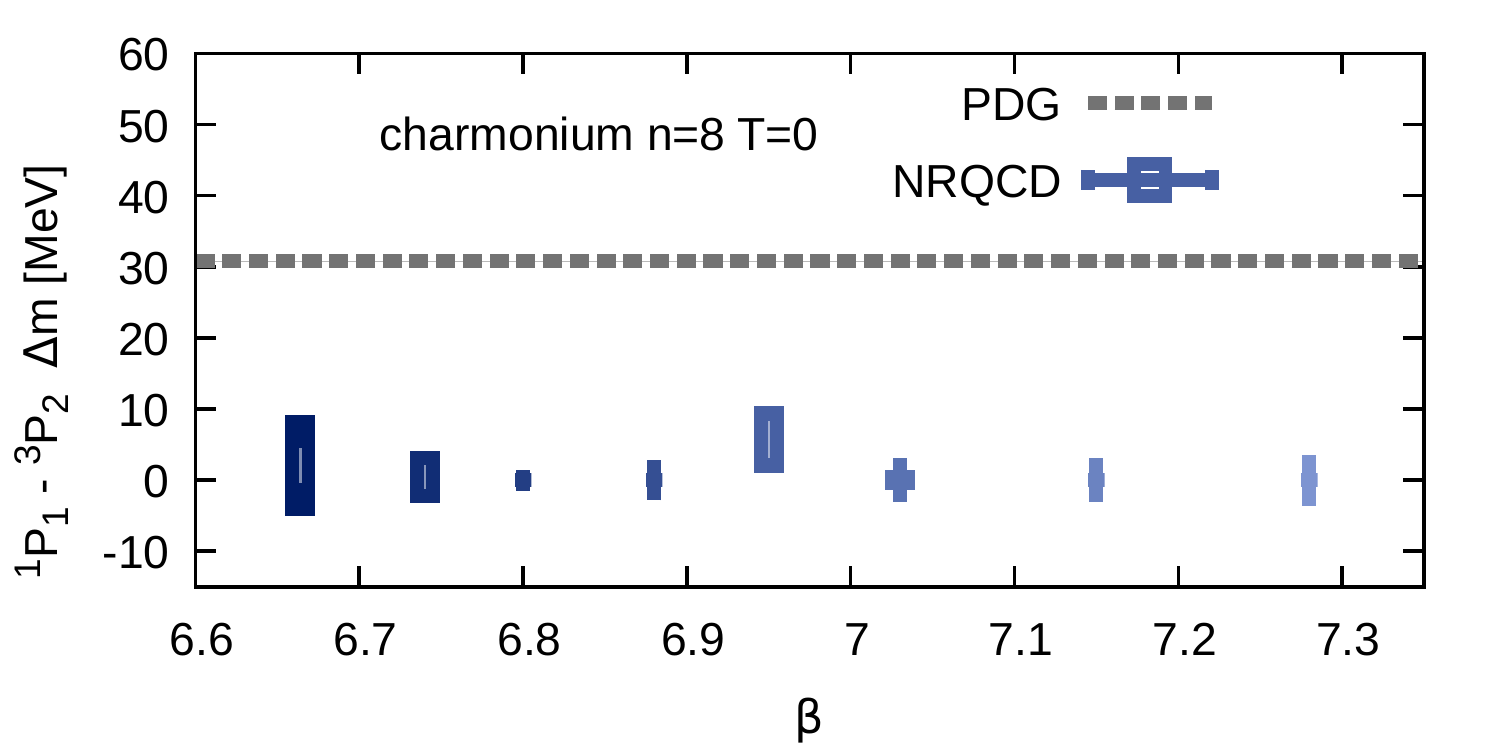}
   \includegraphics[scale=0.30,trim=0cm 0.99cm 0cm 0cm, clip=true]{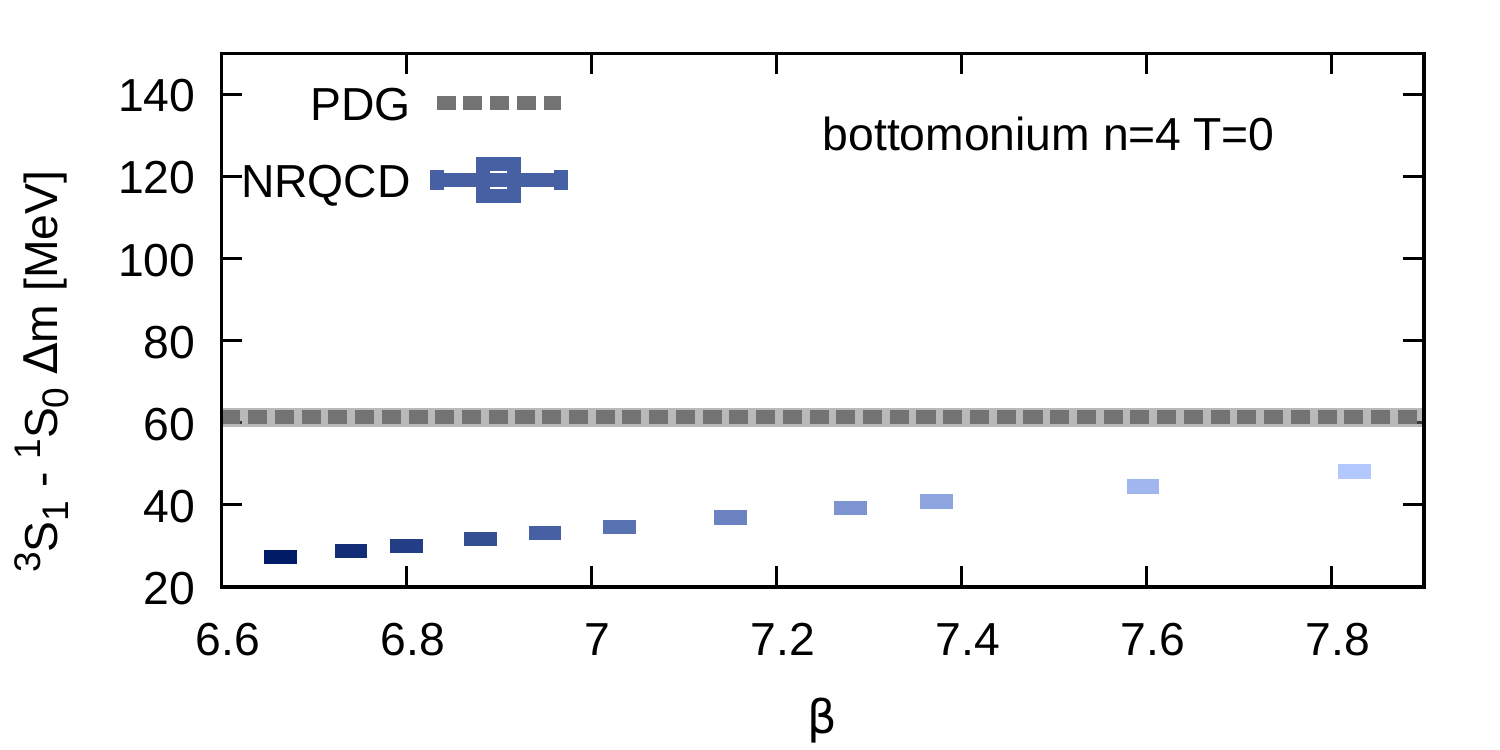}
   \includegraphics[scale=0.30]{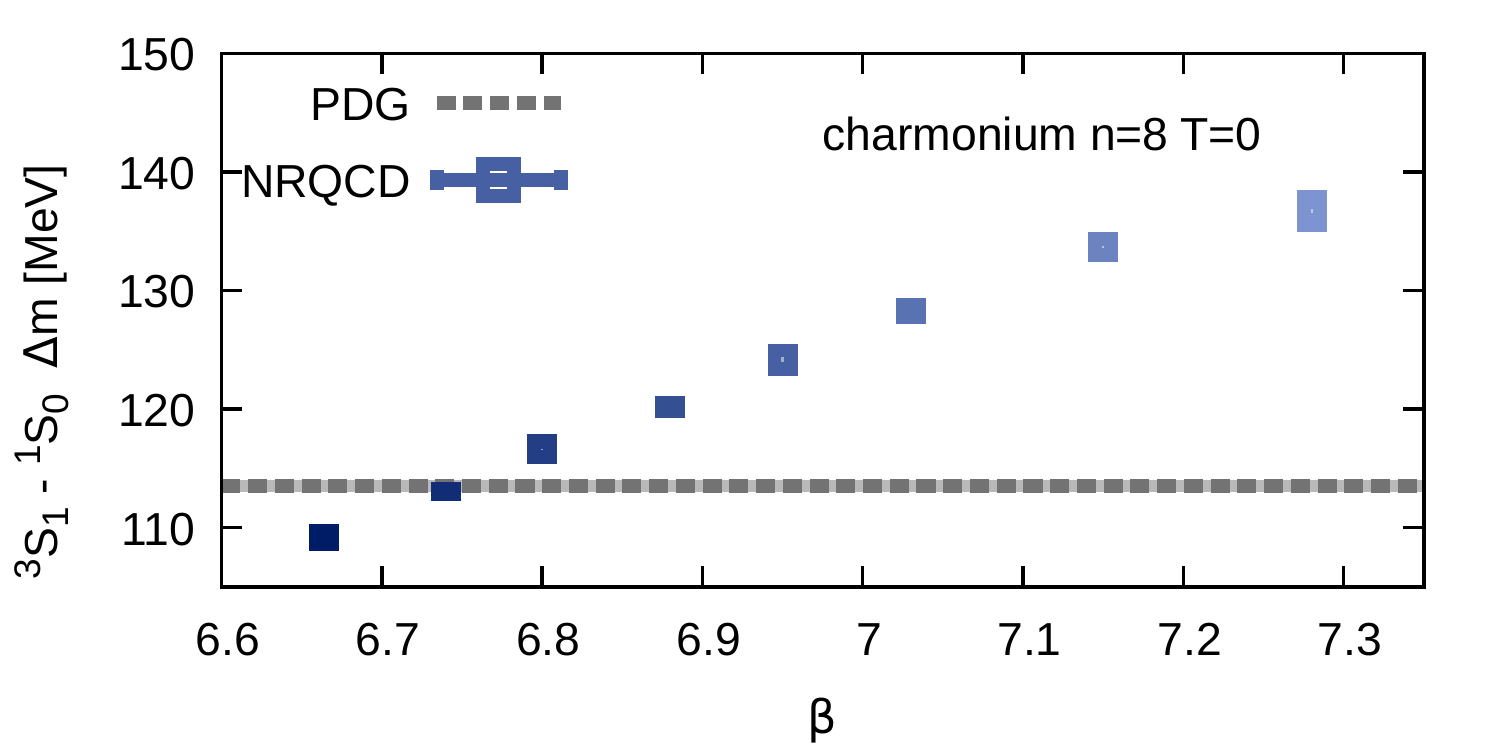}
  \end{center}\vspace{-0.9cm}
 \caption{Mass splittings 
 }\label{Fig:MassSplit}\vspace{-2.cm}
\end{wrapfigure}The resulting smoothing efficiently removes ringing but comes with its own hyperparameter $\kappa$, which we need to set self-consistently. We will use the latter method in the spirit of a ``low gain- low noise'' detector for spectral functions to determine whether a structure in the reconstruction is a genuine bound state feature and if so deploy the ``high gain - high noise'' variant to extract e.g.\ spectral positions.

The $\kappa$ parameter may be selected using prior information in the form of the analytically known non-interacting spectral functions. Reconstructing these from non-interacting lattice correlators, discretized along $N_\tau=12$ Euclidean points shows (Fig.\ref{Fig:T0Prep} bottom) that with $\kappa=1$ no more ringing artifacts remain and an accurate reproduction of the frequency range up to $2$GeV (in terms of $E_{\rm bind}$) can be obtained. In contrast the smoothing properties of the MEM are encoded implicitly in the choice of basis functions and thus vary with the number of available datapoints.\vspace{-0.3cm}

\section{Preparations at T=0}

 As our NRQCD setup is geared to the study of $T>0$ quarkonium properties we forego improvements applicable at $T=0$, such as e.g.\ extended operators. To ascertain how accurate our simulations are in that case, we compute different mass splittings as shown in Fig.\ref{Fig:MassSplit}. Consider a potential based computation: the spin averaged mass split between P-wave and S-wave would only depend on the central potential and thus should be most easily reproduced by NRQCD. The good agreement of the numerical results (top 2 panels, blue points) at different lattice couplings (larger $\beta$ means smaller $a$) compared to the experimental value (gray dashed) confirms this. A more difficult splitting is the $^3P_1-^3P_2$ one (center 2 panels) for which NRQCD must reproduce the physics of the spin-orbit coupling.
 
 \begin{wrapfigure}{r}{0.3\textwidth}\vspace{-1.1cm}
  \begin{center}
   \includegraphics[scale=0.28]{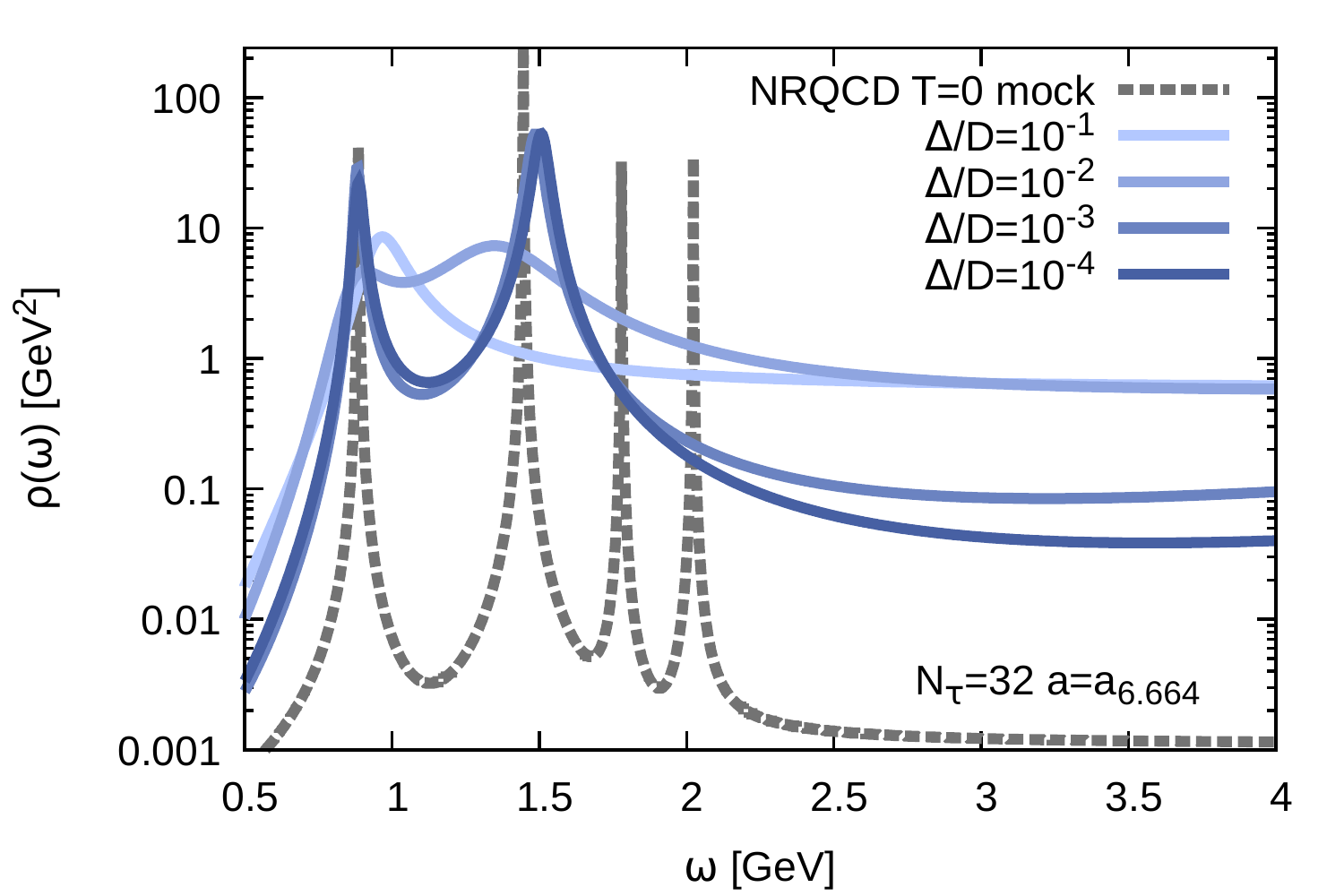}
 \includegraphics[scale=0.28]{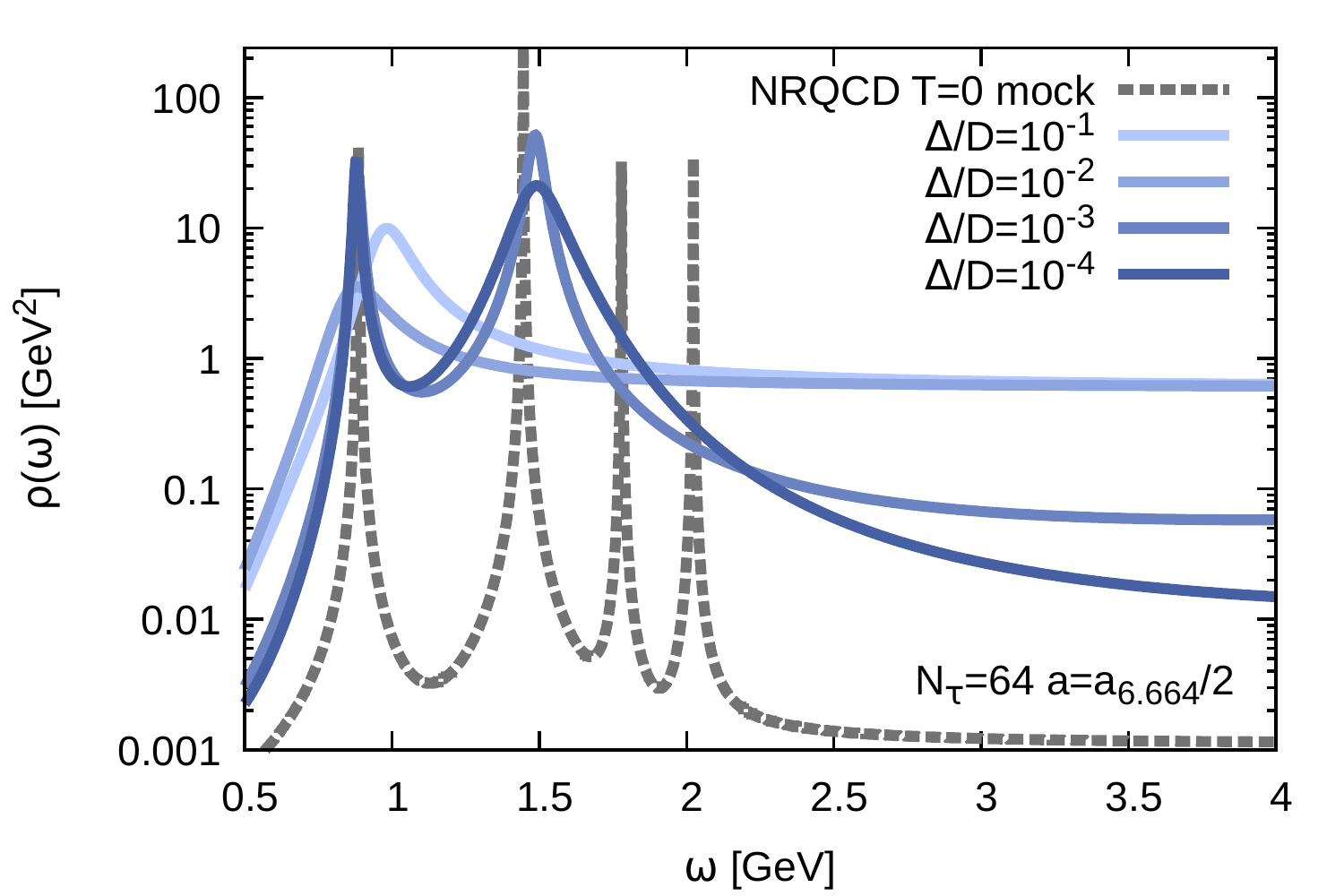}
 \includegraphics[scale=0.28]{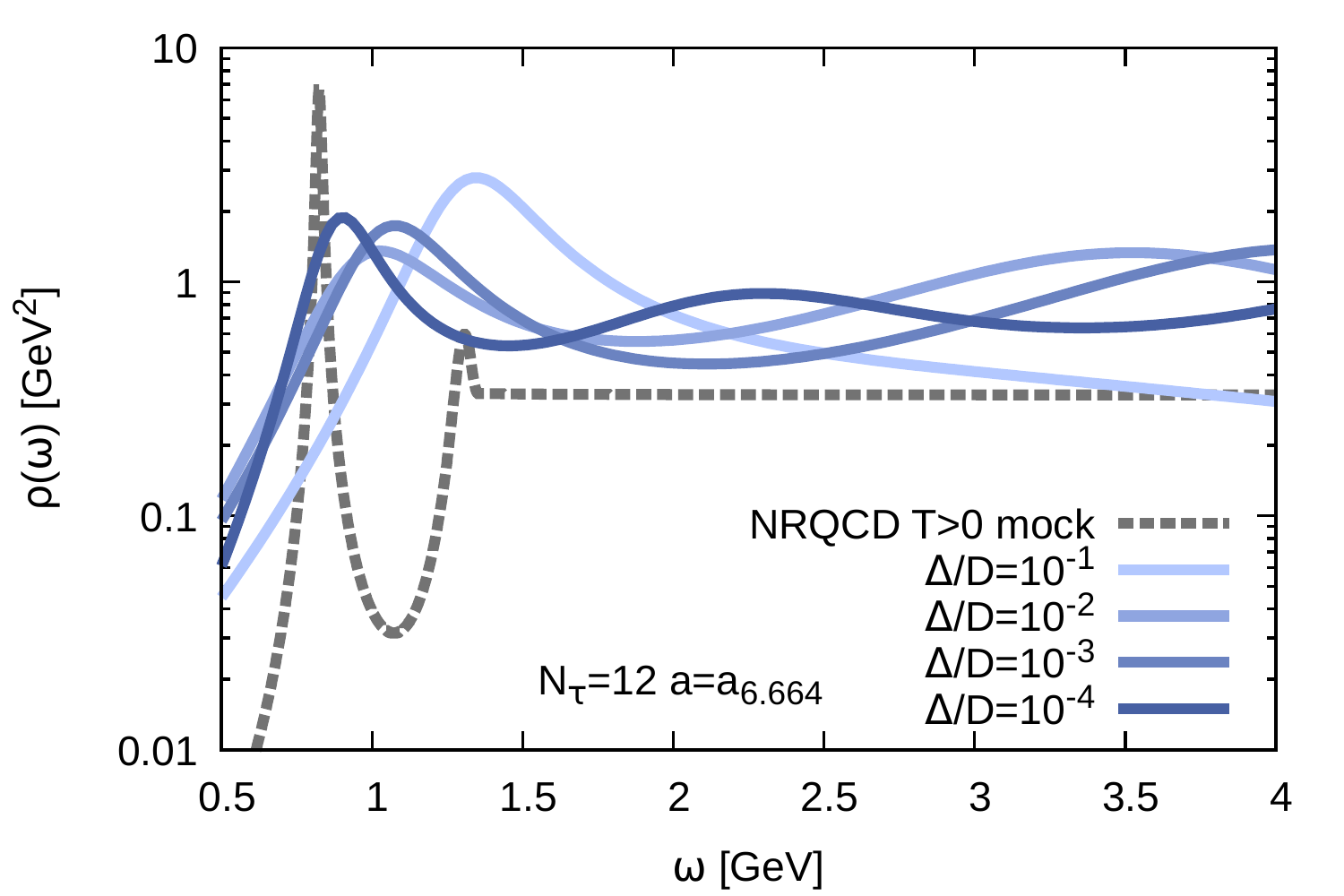}
 \includegraphics[scale=0.28]{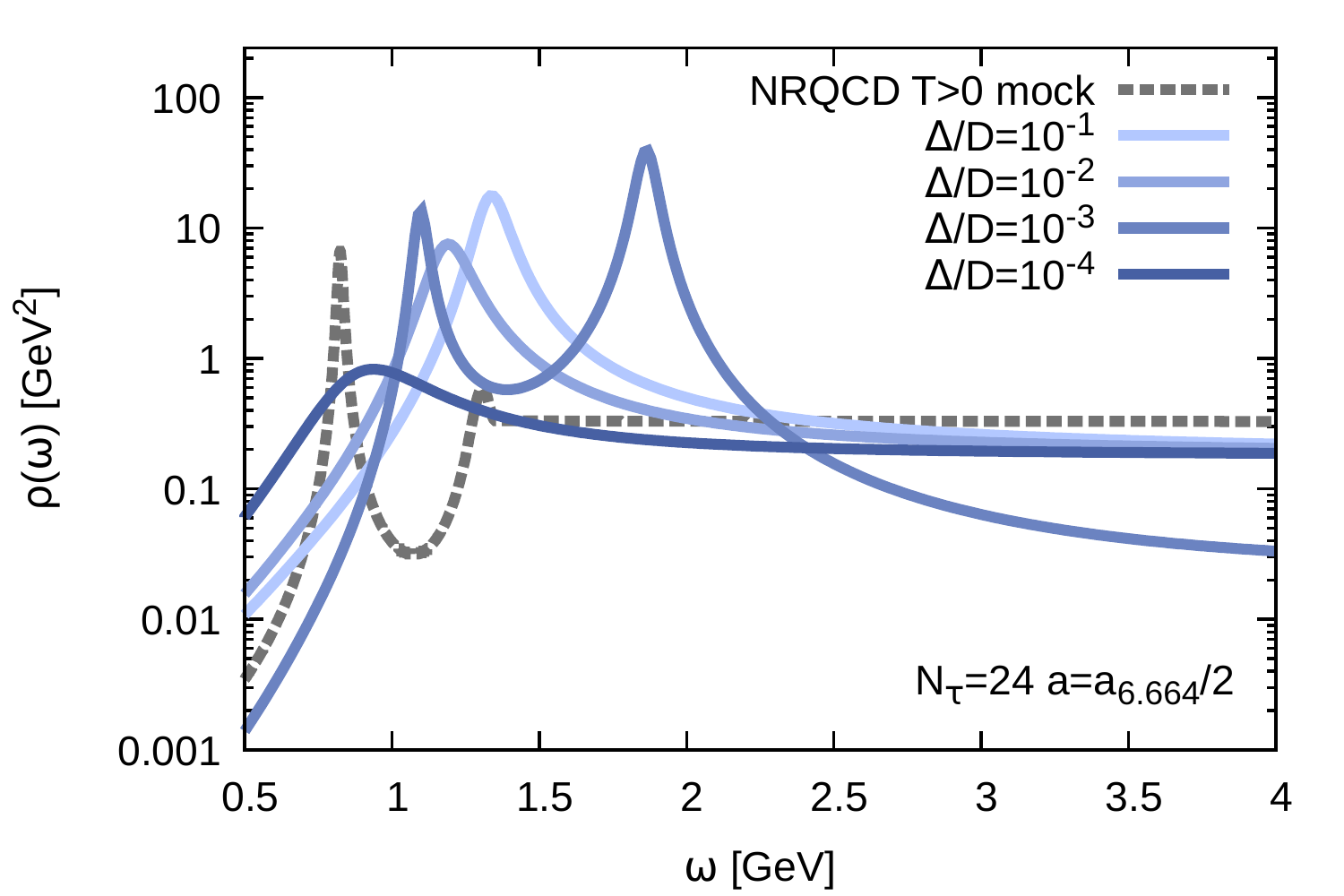}
  \end{center}\vspace{-0.9cm}
 \caption{Mock data test: (top 2) $T=0$ spectrum at different lattice spacings (bottom 2) $T>0$ spectrum at different lattice spacings}\label{Fig:MockTsts}\vspace{-0.5cm}
\end{wrapfigure}
\noindent While for $b\bar b$ we stay close to the physical value within sizable uncertainties, we miss the value in $c \bar c$. Note that the maximum deviation is around $30$MeV. The most difficult splitting known is the S-wave hyperfine splitting, which requires ${\cal O}(v^6)$ and nontrivial radiative corrections to be well reproduced in NRQCD. The bottom two panels show that there is a deviation in our results but always smaller than $35$MeV. We learn that all of our results do carry at most an implicit $35$MeV systematic uncertainty band, leading to an acceptable $T=0$ reference but not being competitive with dedicated $T=0$ NRQCD studies.

The next question we address is how much information is actually stored in the $T=0$ correlators, i.e. accessible \begin{figure}[b]\vspace{-0.5cm}
\centering
 \includegraphics[scale=0.32]{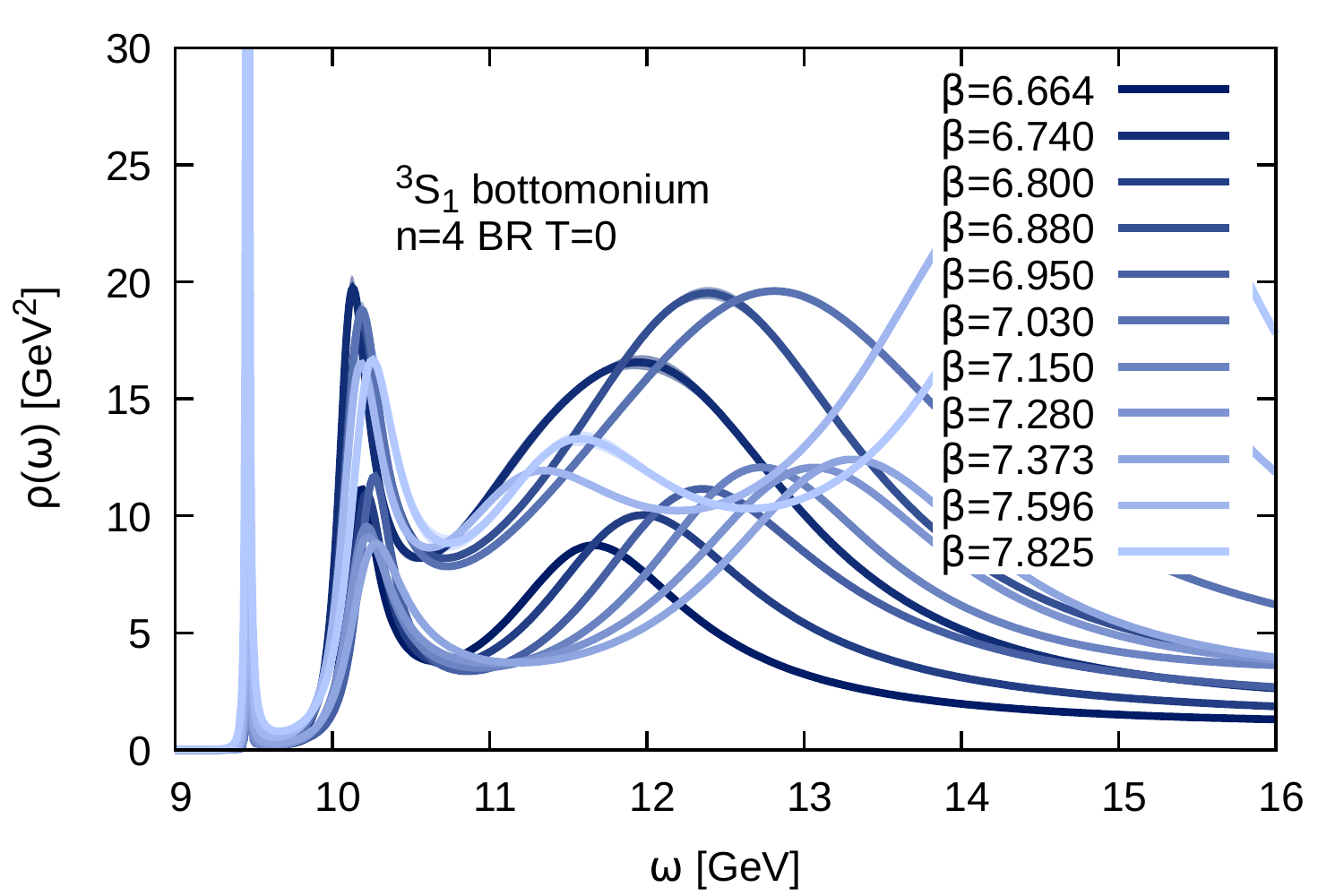}
 \includegraphics[scale=0.32]{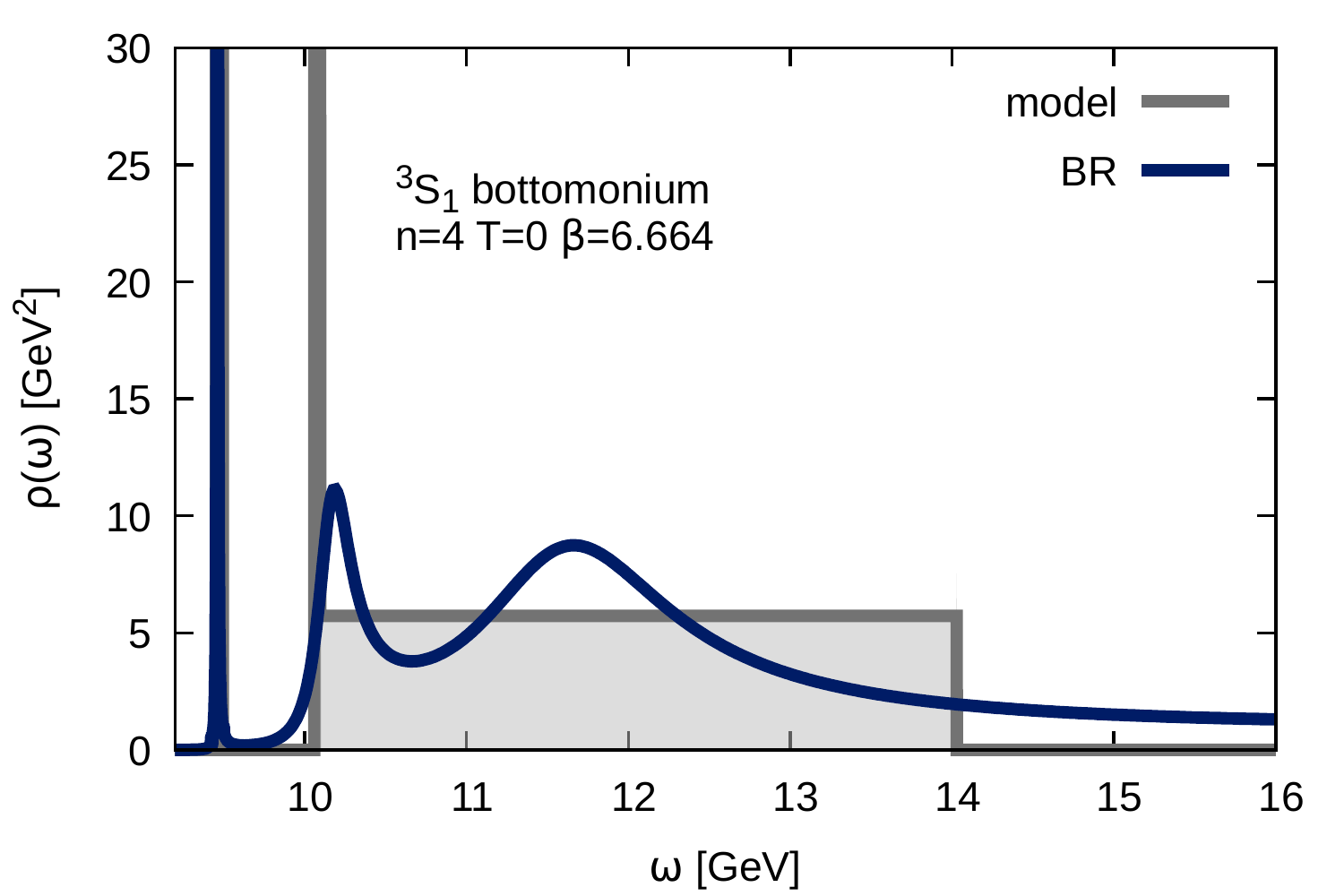}
 \includegraphics[scale=0.32]{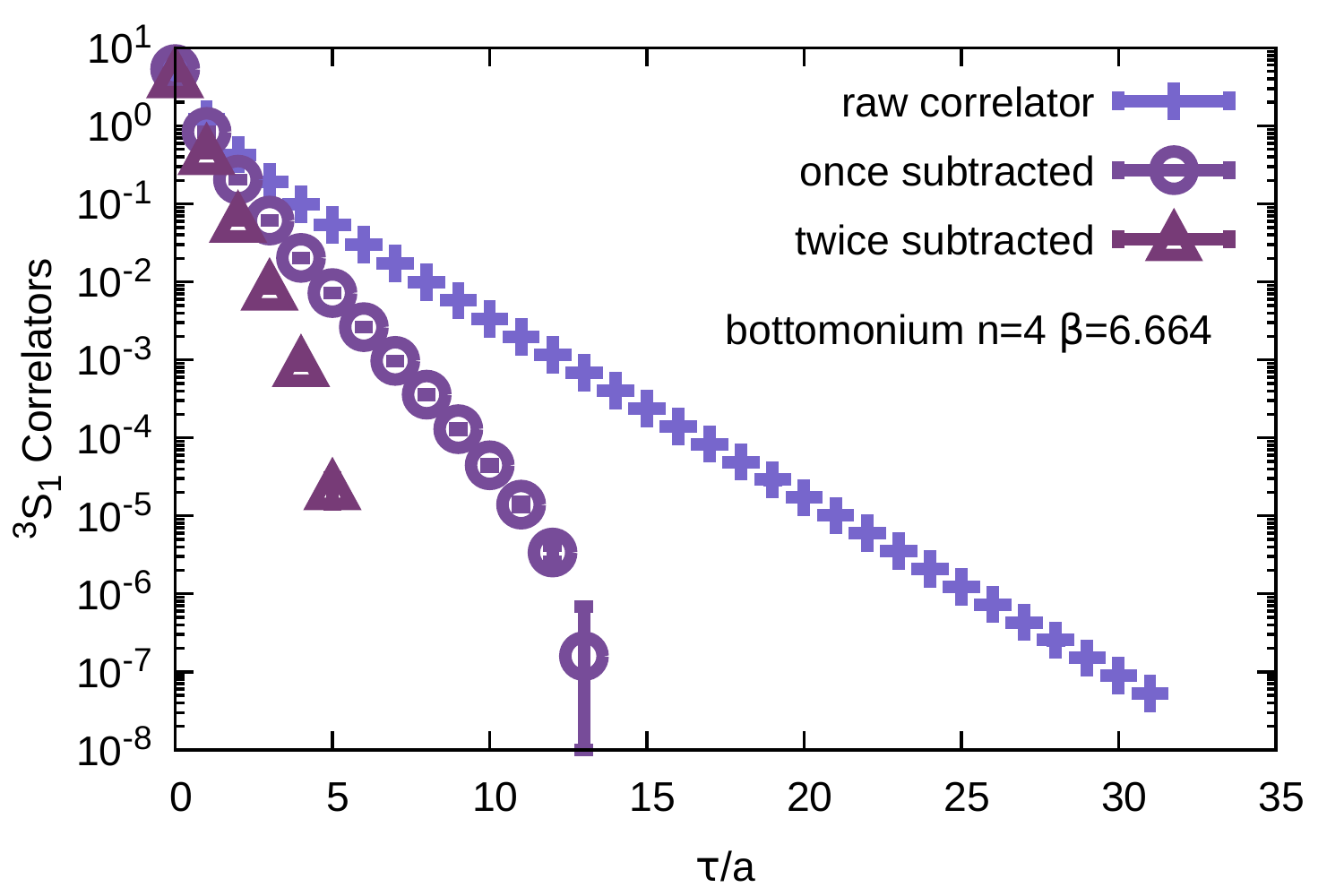}
 \caption{(left) $T=0$ BR spectral reconstruction of $^3S_1$ bottomonium, (center) simplified peak+box model fit to the $T=0$ correlator (right) subtracted $T=0$ correlation function.}\label{Fig:InfoContent}
\end{figure}to a spectral reconstruction. Fig.\ref{Fig:InfoContent} shows the reconstruction of the $\Upsilon$ channel on the left for different lattice spacings. In the center plot we present an alternative interpretation of the correlator data using a simple two-peak and box model, which already manages to reproduce the input within its uncertainty. And indeed after subtracting the lowest two peaks from the correlator (right) we see that only three convex points remain from which the whole excited states and continuum regime needs to be extracted. With current simulation data it appears that we are not just limited by the Bayesian reconstruction but by the information content itself.

To understand what it takes to improve on these results we carried out mock data tests encoding first a $T=0$ like spectrum without continuum in a Euclidean correlator discretized with the same lattice spacing as $\beta=6.664$ at $N_\tau=32$ and with half that spacing at $N_\tau=64$. The sobering result of comparing the corresponding reconstruction (Fig.\ref{Fig:MockTsts} top 2 panels) is that going closer to the continuum limit will not significantly improve the outcome, since the physical extent of the Euclidean lattice remains the same. We have repeated the same test with a $T>0$ like mock spectrum including a continuum part (bottom 2 panels), which tells us that better resolution in $\tau$ will help us better understand the continuum structure, while not improving significantly the bound state reconstruction. This outcome motives increased efforts in bringing lattice QCD simulations on anisotropic lattices to the same level of realism as available today on isotropic ones.

\begin{wrapfigure}{r}{0.61\textwidth}\vspace{-0.6cm}
  \centering
 \includegraphics[scale=0.30]{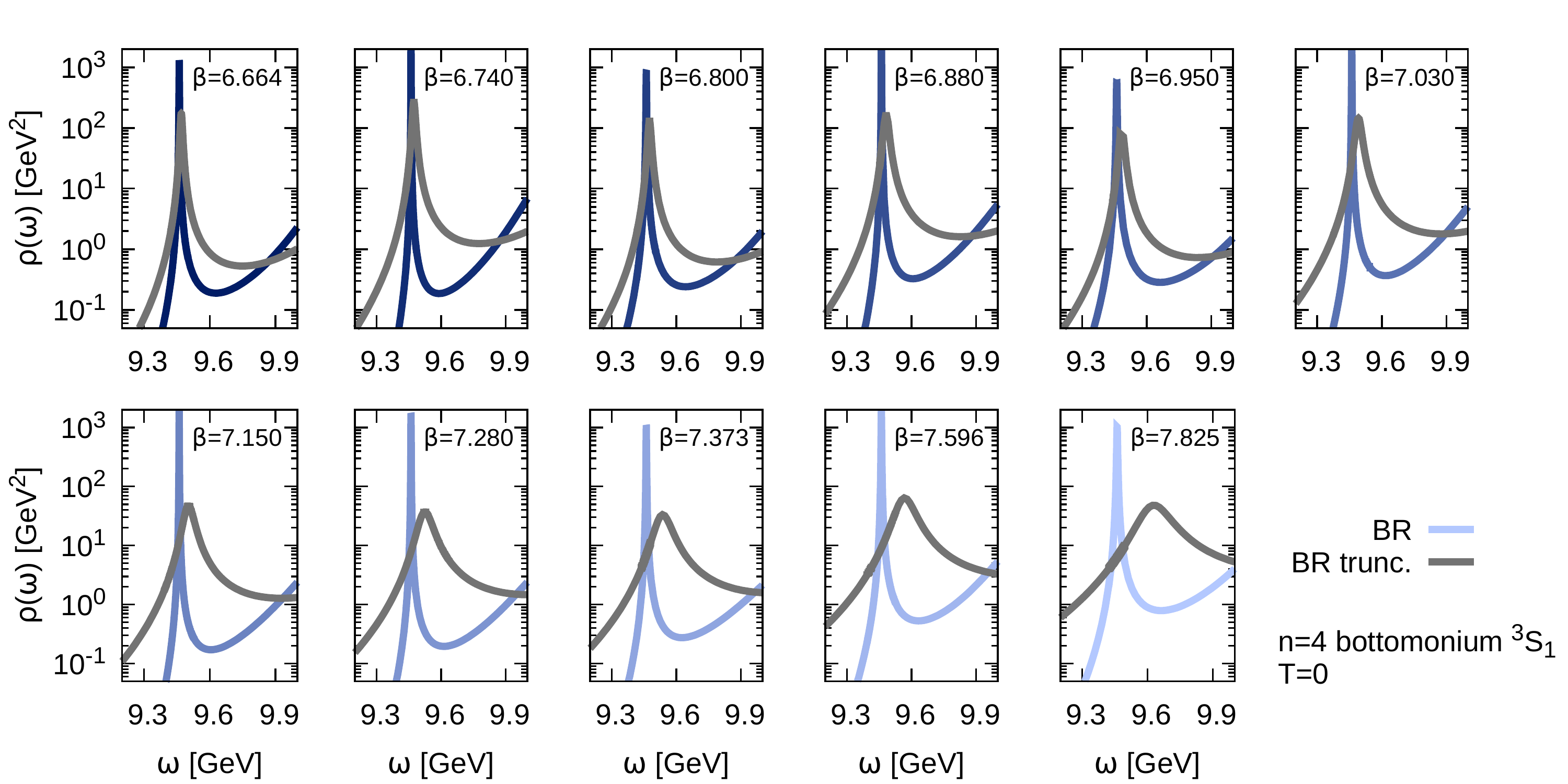}\vspace{-0.4cm}
 \caption{Comparison between reconstructions on the full $T=0$ correlator colored solid) and truncated correlators (gray)}  
\label{Fig:FullVsTrunc}\vspace{-0.2cm}
\end{wrapfigure}Before investigating quarkonium in-medium modification, let us consider another systematic uncertainty, which will be crucial to interpret in-medium effects. The main difference between $T=0$ and $T>0$ is the much smaller number of available input correlator points. While the BR method is able to reproduce peak positions accurately with $N_\tau=32-64$ points we need to quantify how the reconstruction degrades for smaller $N_\tau=12$. To this end we truncate the $T=0$ correlator sets to the same number of points available at $T>0$ and repeat the reconstruction. Fig.\ref{Fig:FullVsTrunc} shows the outcome with the original reconstruction as colored solid line and the one from truncated data as gray lines. 

\begin{wrapfigure}{r}{0.31\textwidth}\vspace{-1.cm}
  \begin{center}
   \includegraphics[scale=0.32]{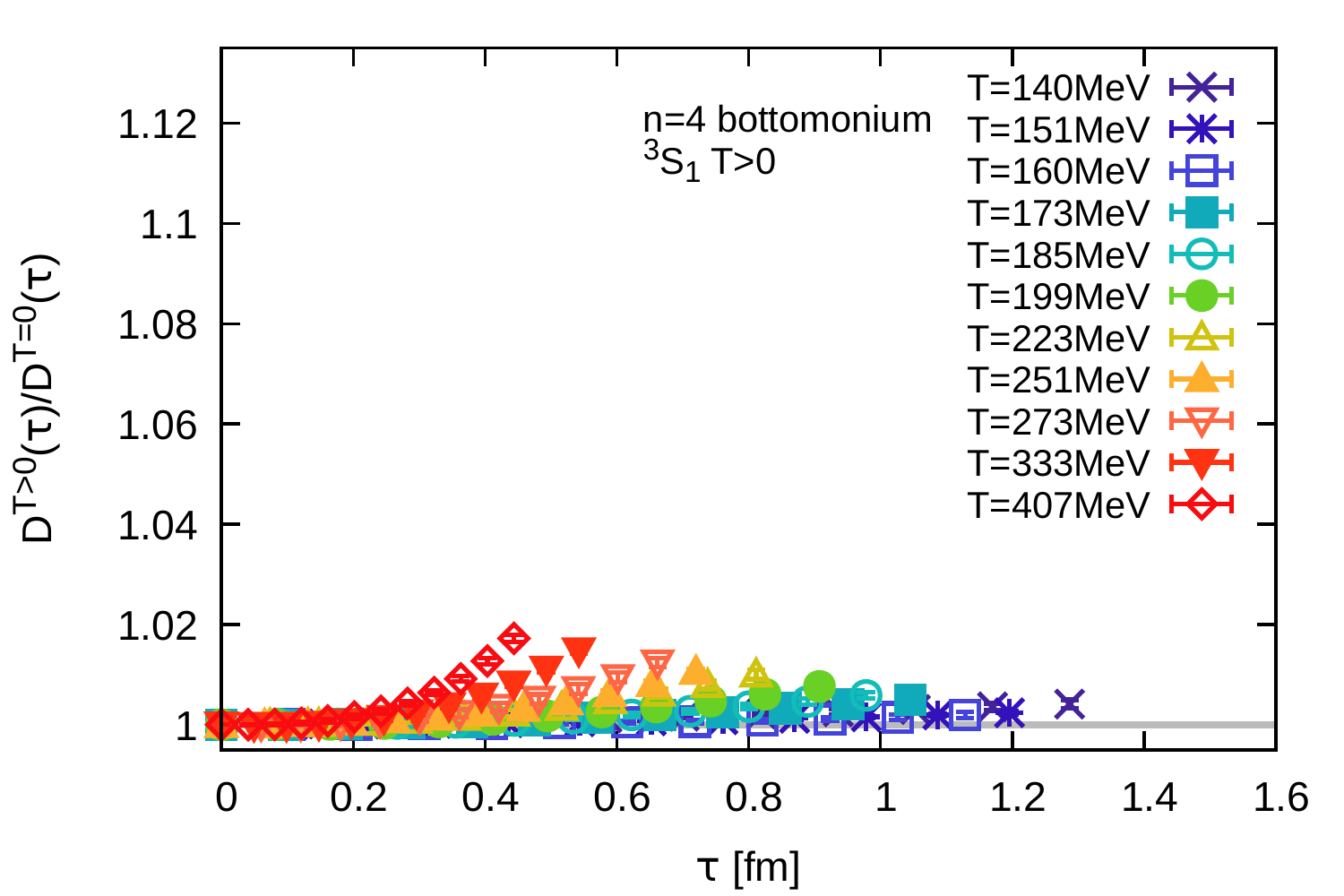}
 \includegraphics[scale=0.32]{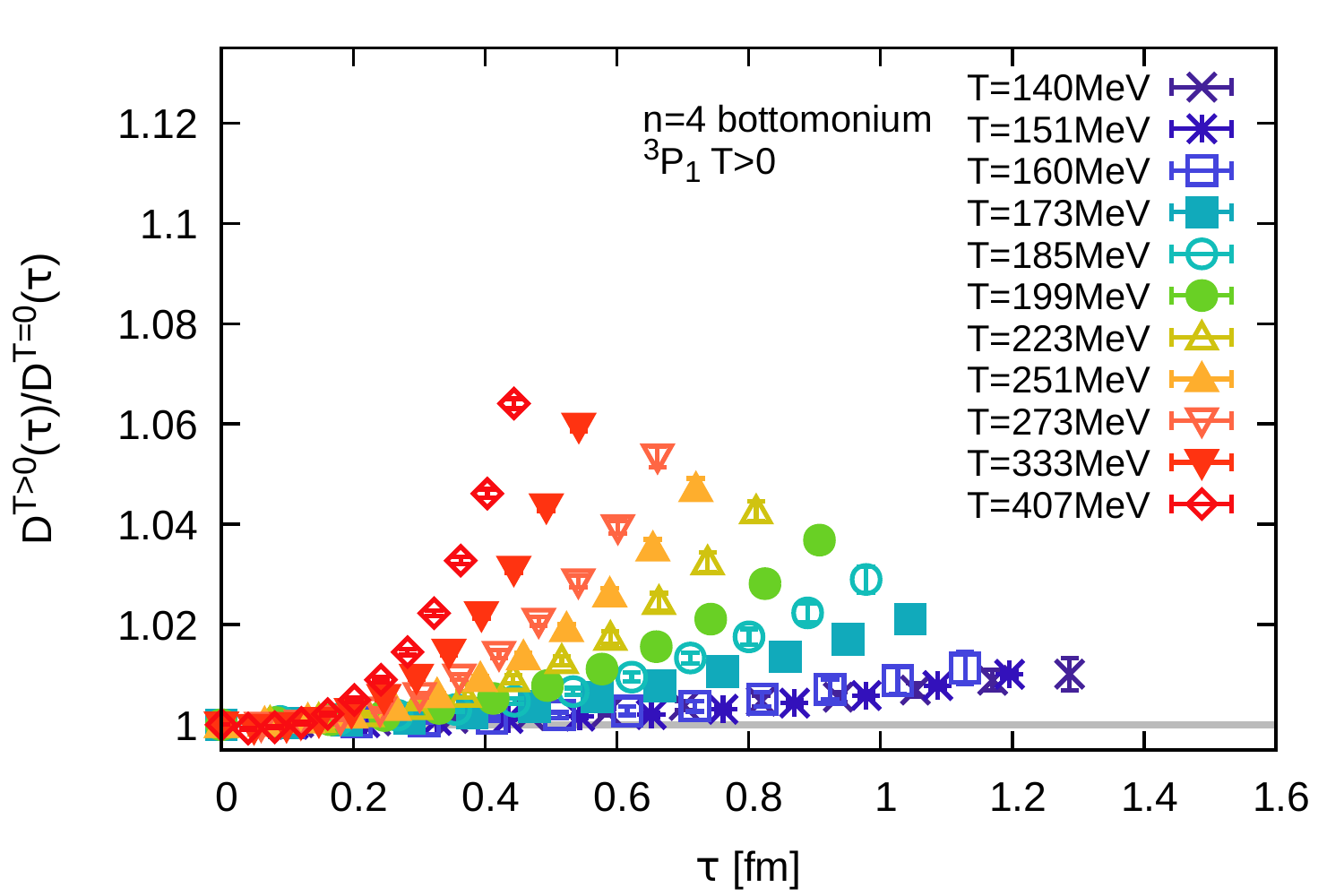}
 \includegraphics[scale=0.32]{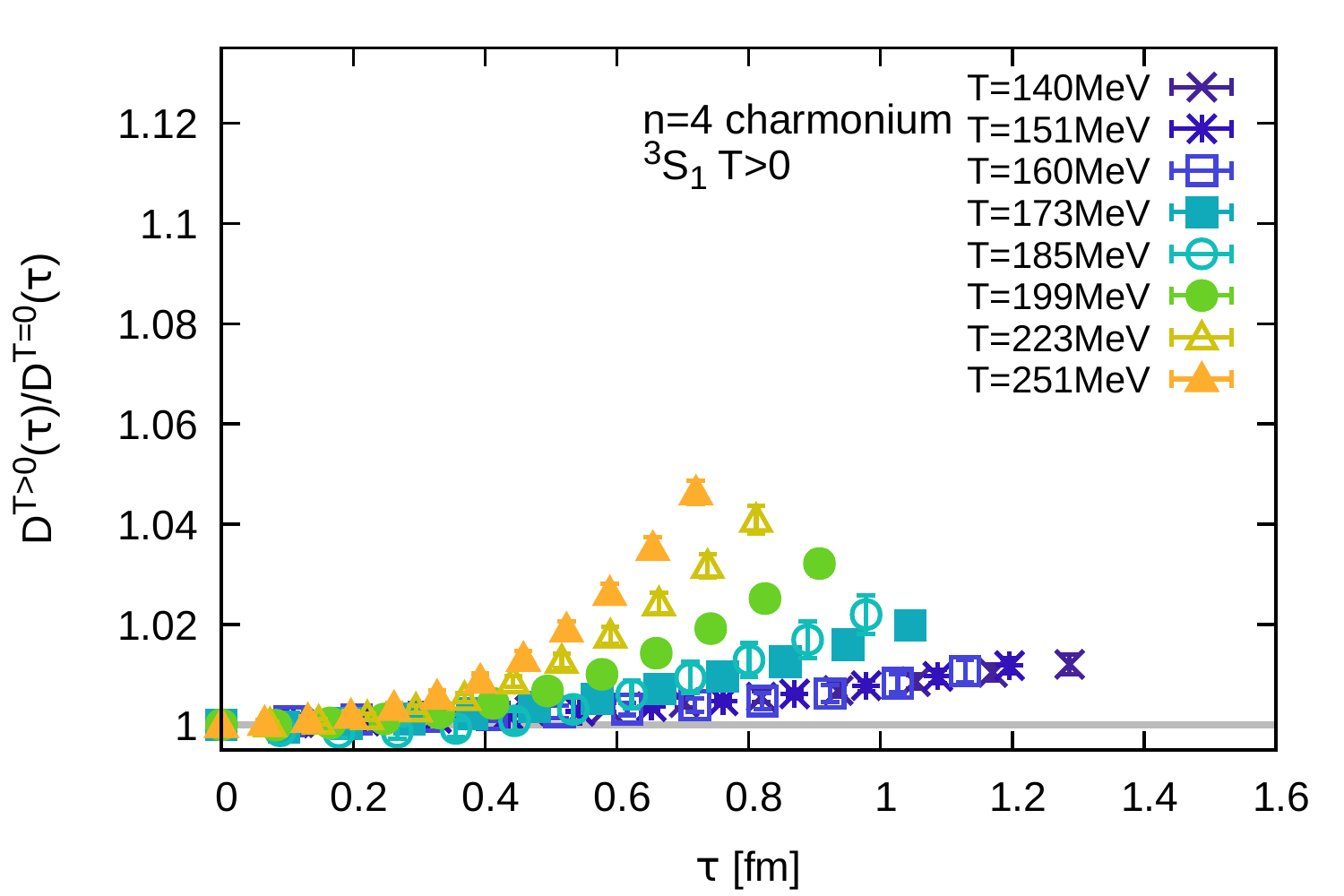}
 \includegraphics[scale=0.32]{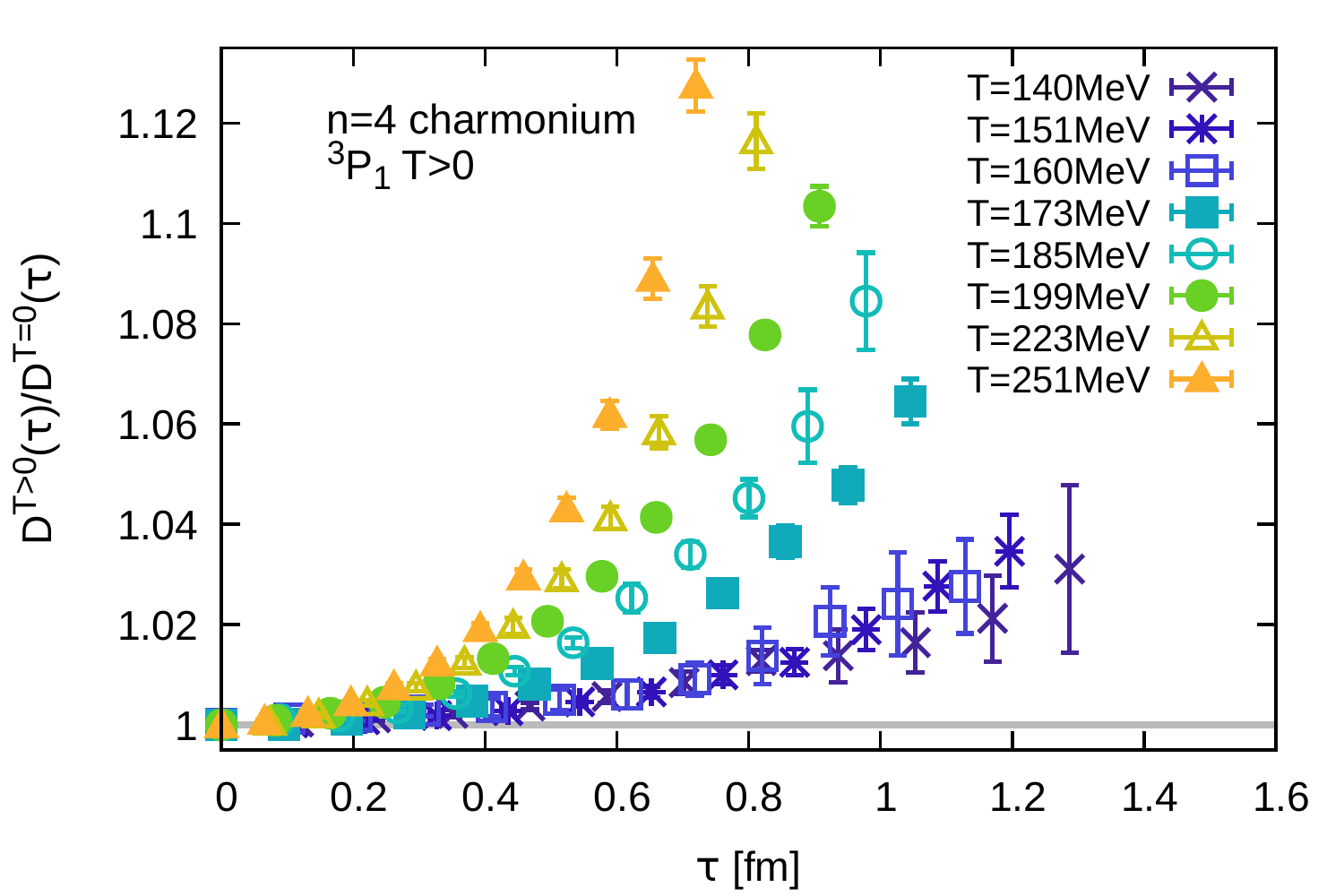}
  \end{center}\vspace{-0.9cm}
 \caption{ Correlator ratios $T>0/T=0$ (top) $\Upsilon$ (2nd top) $\chi_{b1}$ (2nd bottom) $J/\Psi$ (bottom) $\chi_{c1}$ 
 }\label{Fig:CorrRatios}\vspace{-1.6cm}
\end{wrapfigure}
At $\beta=6.664$ where NRQCD works best the difference is very small but at the finer lattice spacings we can clearly see that an artificial shift to higher frequencies and an artificial broadening occurs. We will keep these systematic method artifacts in mind when considering in-medium masses in the next section.\vspace{-0.62cm}

\section{$T>0$ results}

The study of in-medium correlation functions by themselves reveals vital information about the overall in-medium modification of heavy-quarkonium. In Fig.\ref{Fig:CorrRatios} we plot the ratios of the $T>0$ and $T=0$ correlators using the same scale. Already in the hadronic phase we observe deviations form unity hinting at in-medium modification. In the QGP a characteristic upward bend appears, which, as can be learned from comparison with potential based computations \cite{Burnier:2015tda,Burnier:2016kqm}, is indicative of the ground state peak shifting to lower masses and broadening.

Now that we have included charmonium, we can corroborate a picture of a hierarchical in-medium modification of quarkonium ordered by the vacuum binding energy. Indeed for $\Upsilon$ with $E^{T=0}_{\rm bind}=1.1$GeV at $T=407$MeV we have a mere $1.75\%$ deviation (top) while $\chi_{b1}$ with $E^{T=0}_{\rm bind}\approx 640$MeV already shows $6.5\%$ (second from top). While for charmonium we restrict ourselves to lower temperatures, compared at $T=251$MeV we find that $J/\Psi$ with a very similar $E^{T=0}_{\rm bind}\approx 640$MeV as $\chi_{b1}$ shows the same $5\%$ deviation (see the orange points). The much more weakly bound $\chi_{c1}$ with $E^{T=0}_{\rm bind}\approx 200$MeV exhibits around $13\%$ deviation at $T=251$MeV. I.e. states that are less deeply bound and hence more spatially extended are more easily affected by the medium.

\begin{wrapfigure}{r}{0.64\textwidth}\vspace{-0.8cm}
\centering
 \includegraphics[scale=0.44,trim=0cm 10.5cm 10cm 0cm, clip=true]{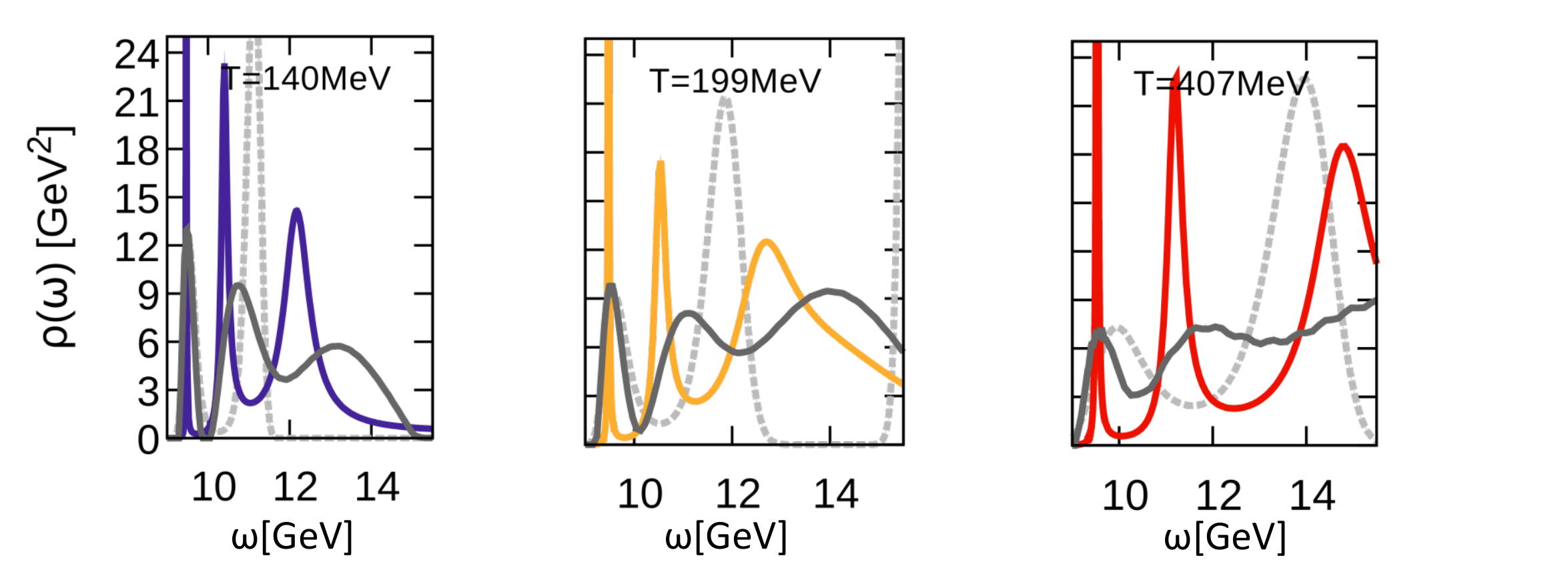}
 \includegraphics[scale=0.43,trim=0cm 9.5cm 10cm 0cm, clip=true]{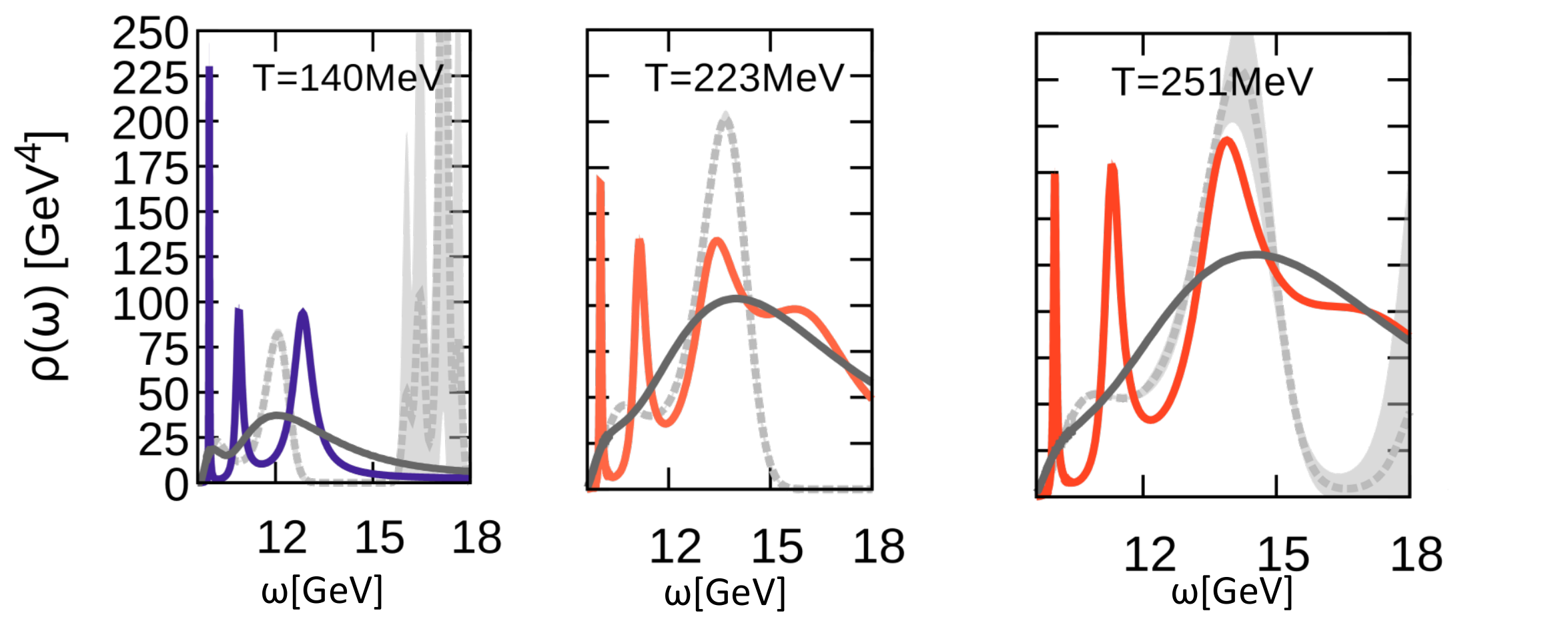}\vspace{-0.3cm}
\caption{Reconstructions of $T>0$ bottomonium spectral functions (top) S-wave (bottom) P-wave using the BR (color solid), smooth BR (gray solid) and MEM (gray dashed). }  \label{Fig:BottomSpecs}\vspace{-0.3cm}
\end{wrapfigure}
\noindent It is important to stress that this finding does not immediately translate into a sequential suppression pattern of quarkonium in HICs.

\begin{wrapfigure}{r}{0.65\textwidth}\vspace{-0.5cm}
\centering
 \includegraphics[scale=0.4,trim=0cm 11cm 9cm 0cm, clip=true]{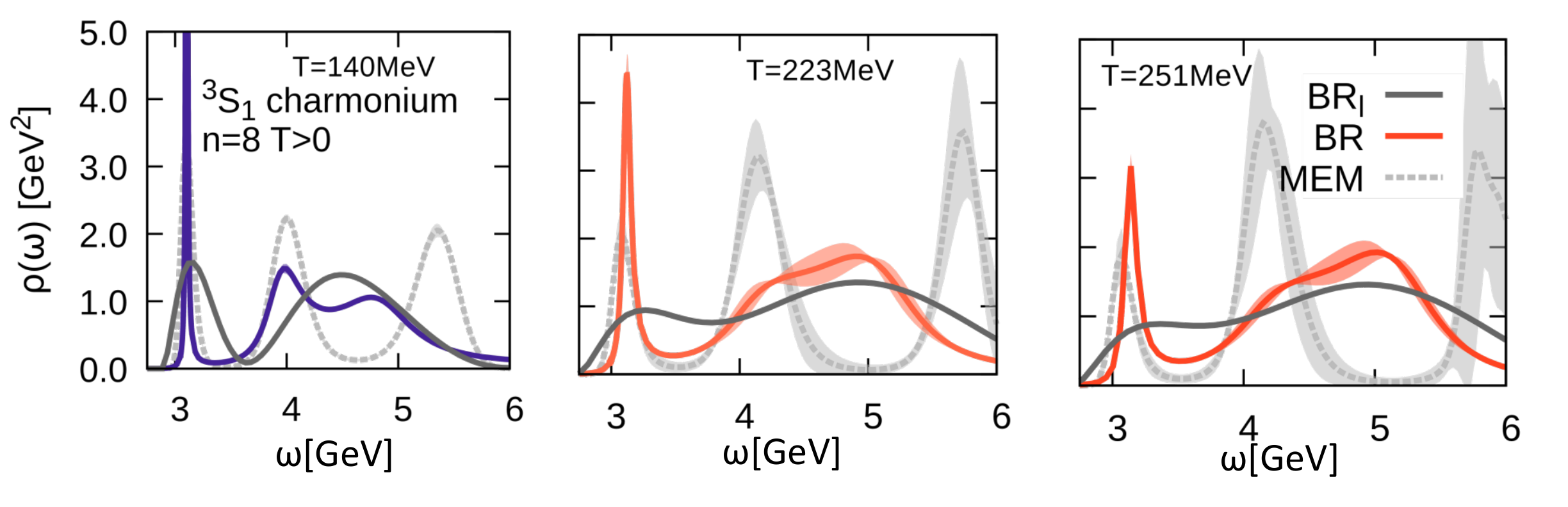}
 \includegraphics[scale=0.41,trim=0cm 11cm 9cm 0cm, clip=true]{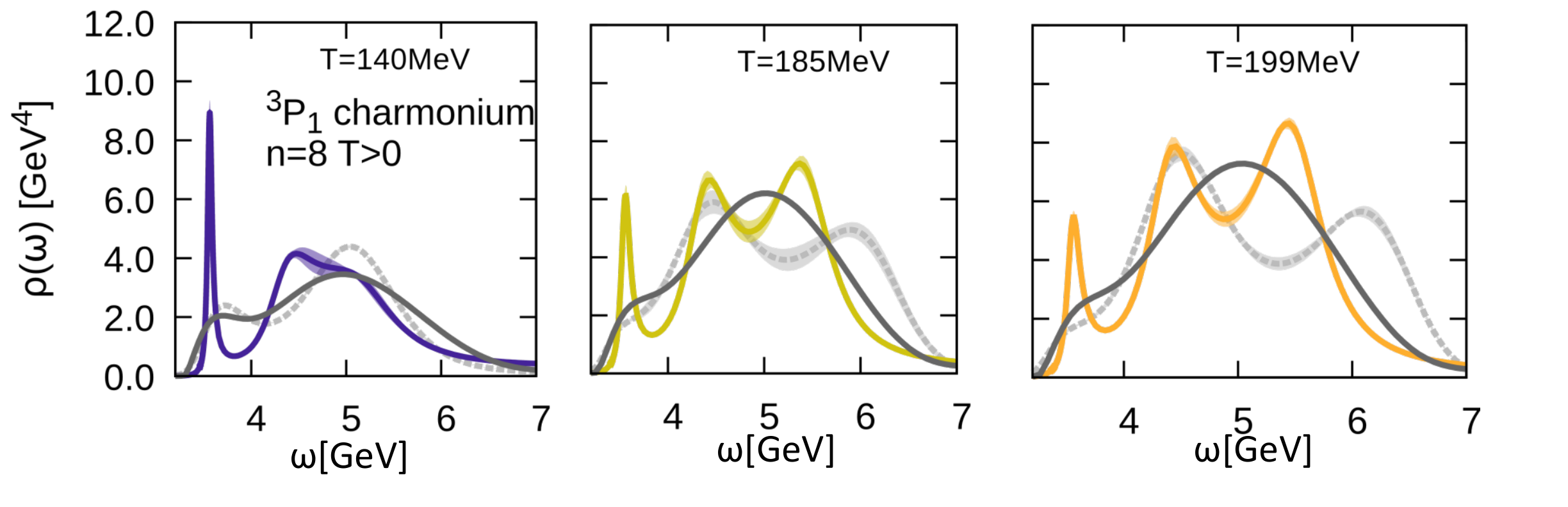}\vspace{-0.3cm}
\caption{Reconstructions of in-medium charmonium spectral functions (top) S-wave (bottom) P-wave 
}  \label{Fig:CharmSpecs}\vspace{-0.5cm}
\end{wrapfigure}

 We continue by carrying out spectral reconstructions on the in-medium correlation functions using the MEM, the standard BR method and the smooth BR. The results for bottomonium and charmonium at three relevant temperatures are shown in Fig.\ref{Fig:BottomSpecs} and Fig.\ref{Fig:CharmSpecs} respectively. Access to the spectral function allows us to estimate the melting temperature of the ground state. Note that the phenomenological concept of melting temperature is not uniquely defined, as the states are expected to monotonously broaden and smoothly merge with the continuum. 
 
 A common criterion in the literature is to define melting as the temperature at which the in-medium binding energy equals the thermal width of the state. This definition however requires detailed knowledge about the continuum threshold, which we are yet unable to resolve in our study. Therefore we revert to the naive definition deployed also in other lattice studies of declaring melting once no remnant of a ground state peak is present anymore. This is much more challenging from the point of view of reconstructions than a mass determination, since it is related to correctly estimating the area of individual peaks. In practice we find that different reconstruction methods with different systematics uncertainties give slightly different results for the disappearance of the ground state peak. We choose therefore to define a melting region below which all methods provide remnant signals, and above which the majority of methods declare the state melted.
 
 \begin{figure}[b]\vspace{-0.3cm}
\centering
 \includegraphics[scale=0.4]{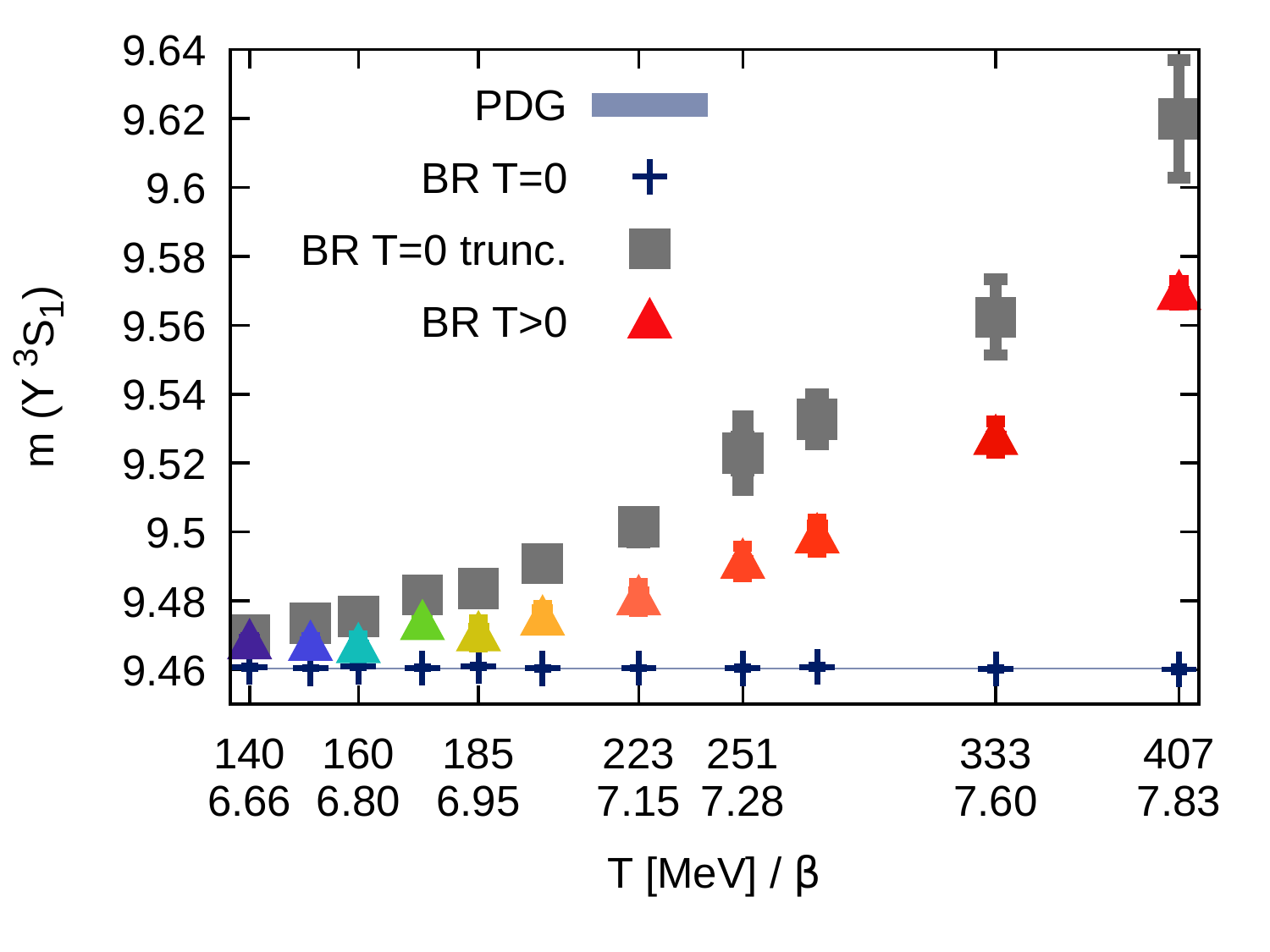}
 \includegraphics[scale=0.4]{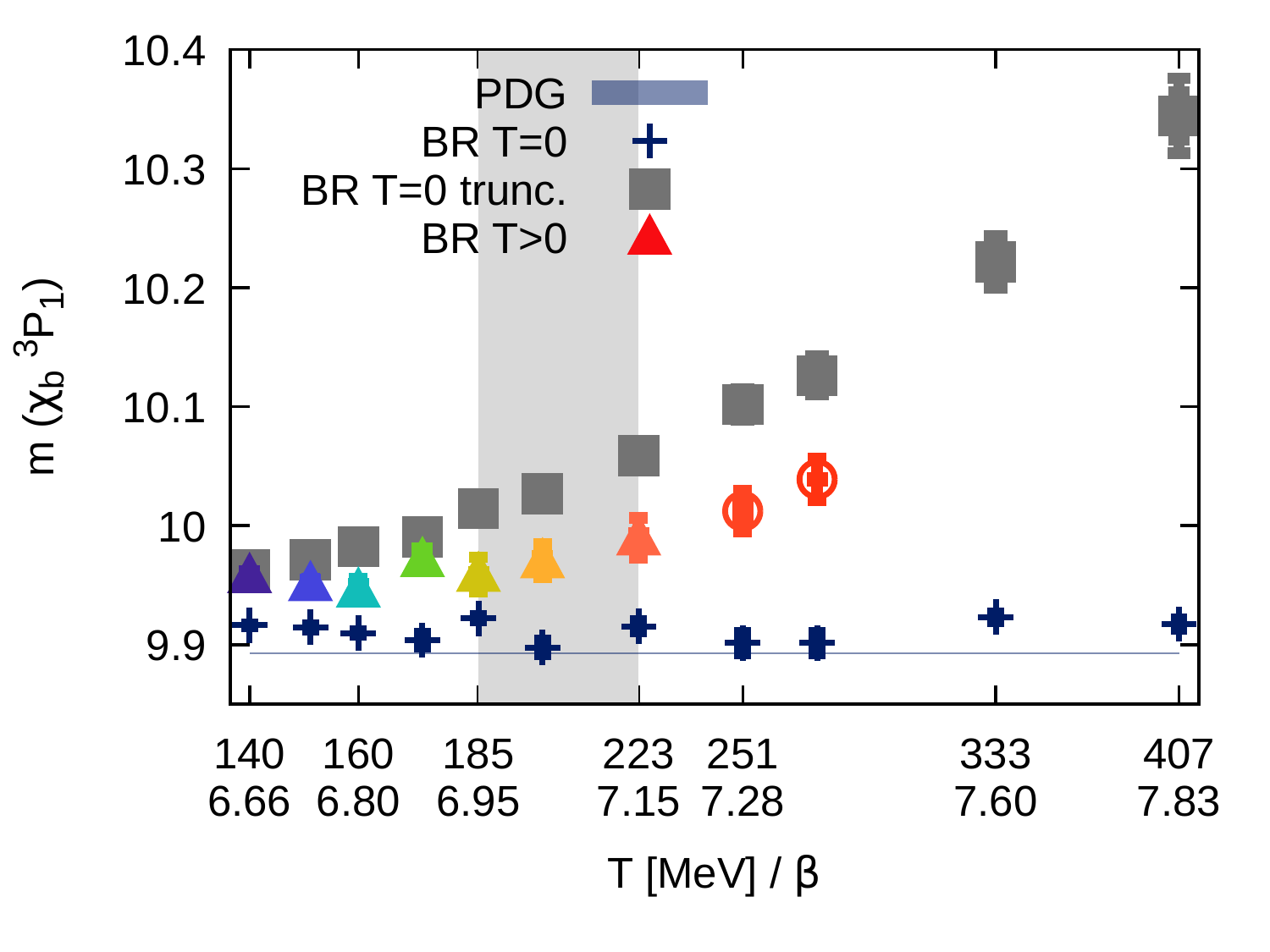}
 \vspace{-0.3cm}
\caption{ (left) $T=0$ NRQCD correlation functions for $^3S_1$ charmonium (right) reconstruction of the non-interacting P-wave spectral functions using three different Bayesian methods}  \label{Fig:BottomShifts}
\end{figure}

Starting with bottomonium S-wave, we find that all three reconstruction methods show a remnant ground state feature at the highest temperature $T=407$MeV. One clearly sees that the standard BR method produces the sharpest ground state peak and the MEM and smooth BR method produce peaks of similar amplitude. At the highest $T$ the BR method appears to suffer from sizable ringing artifacts at higher frequencies that are absent in the smooth BR method. 

In the P-wave bottomonium channel, where previously different melting temperatures were quoted in the literature we can now also compare the three different methods. Here the BR method shows an apparent ground state signal at $T=251$MeV, which we however identify as a ringing artifact, since the smooth BR method and MEM do not show any remnant peak there. Our estimate for the melting region is $T_{\rm melt}\in[185,223]$MeV.  We stress that the MEM result for the ground state peak here is consistent with that obtained in previous studies by the FASTSUM collaboration \cite{Aarts:2014cda}. Our point is that using a single method is not enough to estimate the uncertainties of the reconstruction, leading us to a melting window instead of a specific temperature.
Using a similar strategy for  charmonium in Fig.\ref{Fig:CharmSpecs} reveals melting windows for $J/\Psi$ in $T_{\rm melt}\in[200,210]$MeV and for $\chi_{c1}$ we have $T_{\rm melt}\lessapprox 185$MeV.

The main quantitative result of our study concerns the in-medium mass shifts of the quarkonium ground states. In Fig.\ref{Fig:BottomShifts} we show all necessary ingredients, the blue crosses denote the mass obtained from the $T=0$ spectra extracted from the full correlator data, while the colored points denote the in-medium masses extracted from the corresponding $T>0$ spectra. A first naive visual inspection would lead us to conclude that the in-medium masses are larger than at $T=0$. We however argue that the correct baseline to compare to is given by the masses obtained from spectra reconstructed from $T=0$ correlators truncated to the same Euclidean extent present as at $T>0$. These are the gray squares, which show the artificial shifts already discussed in the previous section. The identification of the proper baseline changes the conclusion profoundly, in that the in-medium mass shift is actually negative, consistent with the behavior of the correlator ratios as well as the predictions of strongly coupled pNRQCD potential based computations. Again the in-medium modification is hierarchically ordered with the vacuum binding energy being stronger for $\chi_{b1}$ than for $\Upsilon$. We would like to stress that the raw in-medium masses obtained here are similar to those found in previous studies by e.g. the FASTSUM collaboration, however the different choice of baseline leads to different conclusions for the in-medium mass shifts.
 
We have presented updated results on the in-medium modification of $b \bar b$ and $c \bar c$ states using lattice NRQCD with significantly increased statistics and a larger $T$ regime. With charm d.o.f. included we confirmed that in-medium modification is hierarchically ordered with the $T=0$ binding energy. Comparing different Bayesian reconstruction methods allowed us to estimate temperature windows for  melting of individual states, showing that previous discrepancies reported in the literature are due to underestimated methods uncertainties. Carefully selecting a proper baseline for $T=0$, we extracted robust estimates for the in-medium mass shifts, which we find to be negative and hierarchically ordered in agreement with strongly coupled pNRQCD.   

S.K. acknowledges funding by NRF grant NRF-2018R1A2A2A05018231, PP by the U.S. DOE contract No.DE-SC001270 and A.R. via the DFG collaborative research center SFB1225 ISOQUANT as well as USQCD computing time grants at the JLAB facility. 
 
\bibliographystyle{abbr}
\bibliography{NRQCD}

\end{document}